# Novel $A_2CrH_6$ (A = Ca, Sr, Ba) hydrides explored by first-principles calculations for hydrogen storage applications

Zakaria El Fatouaki [1, *], El Mustapha Hrida[1], Abderahhim Jabar [2,3], Abdellah Tahiri [4,5], Mohamed Idiri [1]

[1] LCMP, Laboratory of Condensed Matter Physics, Hassan II University, Faculty of Sciences Ben M'Sik, B.P. 7955, Casablanca, Morocco

[2] LMHEP, Faculty of Sciences Aïn Chock, Hassan II University, B.P. 5366 Casablanca, Morocco

[3] LPHE-MS, Science Faculty, Mohammed V University in Rabat, Rabat, Morocco

[4] ISTM, Laboratory Innovation in Sciences, Technologies, and Modeling, Department of Physics, Faculty of Sciences, Chouab Doukkali University, El Jadida, 24000, Morocco

[5] LM2A, Laboratory of Advanced Materials and Applications, Faculty of Sciences Dhar El Mahraz, Sidi Mohamed Ben Abdellah University, B.P.1796, Fez, Atlas, Morocco

*Corresponding author: zakariaelfatouaki5@gmail.com (Z.E.)

**Abstract**

A theoretical study of a number of properties of $A_2CrH_6$ (where A = Ba, Sr, and Ca) hydride perovskites with the Cambridge Serial Total Energy Package (CASTEP). These include structural, hydrogen storage, mechanical, phonon, thermodynamic, electronic, and optical properties. The lattice constants of the compounds studied are in the range from 7.220 Å to 8.082 Å, and they exhibit stable cubic crystal structures. Negative formation energies, elastic constants, phonon dispersion and AIMD simulations testify to their thermodynamic, mechanical, dynamic and thermal stability, respectively. For the perovskite hydrides $Ba_2CrH_6$, $Sr_2CrH_6$ and $Ca_2CrH_6$, the corresponding specific hydrogen storage capacities are 1.82 wt.%, 2.69 wt.%, and 4.37 wt.%, respectively. Among these compounds, $Sr_2CrH_6$ exhibits the lowest applicable hydrogen desorption temperature, at 463.7 K. The electronic bands show remarkable spin activity, demonstrating that the change of $A^{2+}$ cation (where A = Ca, Sr, and Ba) immediately influences the spin polarization and electronic behavior of hydride perovskites. On the basis of the elastic moduli studied, the mechanical behavior determines that $Ca_2CrH_6$ is the strongest material. The present results highlight the potential of $A_2CrH_6$ (A = Ca, Sr, and Ba) perovskite hydrides, in particular $Ca_2CrH_6$, for applications in advanced energy systems and hydrogen storage, as well as for electrical and optoelectronic devices.

# 1. Introduction

The transition to renewable energy sources on a global scale requires the development of reliable storage technologies capable of meeting improve energy efficiency while reducing greenhouse gas emissions [1, 2]. Hydrogen's high energy density, adaptability and lack of environmental impact when burned or converted to energy make it a particularly attractive energy carrier [3]. However, large scale implementation requires high performance, secure and economically viable hydrogen storage systems [4], capable of fast, reversible and precise hydrogen storage and release [5]. A decisive parameter for verifying the efficiency of these devices and ensuring that their energy capacity is adapted to the needs of transport, industrial applications and energy distribution networks is their high gravimetric capacity [6, 7].

The distinctive crystal structure of perovskite hydrides of the $A_2BH_6$ type [8], which promote the absorption and release of hydrogen with good bulk density and high thermal stability, is attracting increasing interest as potential solid materials. The ease with which performance can be improved as a function of the selection of A and B cations in these compounds not only enables hydrogen to be stored reversibly [9], but also provides significant structural flexibility. In our case, the alkaline-earth elements Ca, Sr and Ba occupy site A, while site B is occupied by Cr. As element B is decisive in chemical interactions with hydrogen, and the type of element A has a profound impact on crystal stability and storage capacity, this combination enables thermodynamic and kinetic performance to be optimized [10]. Perovskite hydrides of the $A_2BH_6$ type provide a potential basis for the design of efficient hydrogen storage materials that meet the needs of new generations of energy systems [11, 12]. They are an excellent choice for stationary and on-board storage devices, thanks to their high structural stability and their ability to ensure high-performance adsorption and desorption cycles. Consequently, a detailed study of $A_2BH_6$ perovskite hydrides is fundamental to the establishment of safe, reliable and economically hydrogen storage devices [5].

To make the most of hydrogen's potential, integrate it into today's energy systems and foster the development of a sustainable [13], clean and competitive energy economy, overcoming present and future environmental and energy challenges, it is important to develop and optimize these solid materials [14].

Recently, Amer Almahmoud et al. [11] presents a comprehensive study of compounds $Y_2CoH_6$ (where Y = Mg, Ca, Sr, and Ba) using the CASTEP program [15], examining their structural, thermodynamic, mechanical, electronic and optical properties. The results demonstrate the thermodynamic, dynamic and mechanical stability of this material, highlighting its reliability. In terms of hydrogen storage, Sr$_2$CoH$_6$ is characterized by a particularly low desorption temperature (221.7 K), making it ideal for practical use, while Mg$_2$CoH$_6$ has the highest gravimetric capacity (5.32 wt.%). The prospect of perovskite hydrides Y$_2$CoH$_6$ (where Y = Mg, Ca, Sr, and Ba), and Mg$_2$CoH$_6$ in particular, for hydrogen storage and applications in advanced energy systems, is generally highlighted by this work.

In addition, another theoretical study by Bilal Ahmed et Al. [12] highlights the value of hydrogen as a green power source, and underlines the crucial role of metallic hydrides as efficient storage devices. This study presents the first in-depth look at the structural, optical, electrical and mechanical properties of hydrides Q$_2$FeH$_6$ (where Q = Sr and Ca) using DFT (GGA-PBE). With indirect bandgaps of 1.67 eV for Ca$_2$FeH$_6$ and 1.37 eV for Sr$_2$FeH$_6$, the results indicate the semiconducting nature of both hydride materials and justify their thermodynamic stability. Optical analysis provides information on dielectric constants, absorption and energy loss functions, expanding our knowledge of potential photoelectric properties. Overall, this study opens the way to new optoelectronic applications, while highlighting the potential of Ca$_2$FeH$_6$ and Sr$_2$FeH$_6$ as interesting candidates for hydrogen storage.

In another theoretical study based on DFT by Zakaria EL FATOUAKI et al [16, 17], an extensive analysis of the structural, electrical, mechanical, thermodynamic, ionic diffusion, and thermal properties of perovskite hydrides LiM$_3$H$_8$ (where M = Fe, Ni, Cr, and Mn) has been carried out. The findings point to their remarkable stability and gravimetric storage capacity (spanning 4.2 wt.% and 4.7 wt.%) at desorption temperatures suited to real-life applications. LiNi$_3$H$_8$ and LiMn$_3$H$_8$ show improved ion diffusion performance and ductility, especially for LiNi$_3$H$_8$ with high conductivity, but LiFe$_3$H$_8$ and LiCr$_3$H$_8$ exhibit brittleness and mechanical anisotropy. These findings confirm the applicability of this category of hydrides as promising materials for the efficient storage and transport of hydrogen in next-generation energy systems.

In another study, El Houcine Akarchaou et Al. [18] used DFT calculations to explore the perovskite hydrides A$_2$MgTiH$_6$ (where A = Li, Na, and K) to determine their hydrogen storage capacities, as well as their electronic, optical and mechanical properties. The findings confirm the elastic stability of these three hydride perovskites according to Born and Huang's criteria, and indicate astonishing storage capacity (6.16 wt.% for Li$_2$MgTiH$_6$, 4.64 wt.% for Na$_2$MgTiH$_6$, and 3.72 wt.% for K$_2$MgTiH$_6$). Metallic behavior is described by electronic analysis, and the difference between fragility and ductility is highlighted by mechanical properties. This investigation suggests that $A_2MgTiH_6$ perovskite materials (where A = Li, Na, and K) are adaptable for hydrogen storage applications.

Throughout this manuscript, the structural, hydrogen storage, electronic, phonon, optical, and elastic properties of the new materials $A_2CrH_6$ (A = Ca, Sr, and Ba) are discussed in light of density functional theory (DFT). This type of hydride perovskite is characterized by a wide range of physicochemical properties. The performance of these materials was influenced by the possibility of adding Ba, Sr, and Ca cations to their crystal structure. Furthermore, the development of durable and efficient energy storage systems is based on the system's ability to absorb and release hydrogen, $A_2CrH_6$ (A = Ca, Sr, and Ba) hydrides have demonstrated their usefulness in hydrogen storage systems. Finally, Research into hydrides $A_2CrH_6$ (A = Ca, Sr, and Ba) presents enormous opportunities for development in energy technology. To this day, no theoretical analysis or experimental trials have been carried out to study these specificities. However, these hydrides could be created in laboratories in the relatively near future and used for tasks such as hydrogen storage. The properties of these perovskites are being evaluated using the CASTEP program [15]. These new combinations of hydride perovskites are among the materials that could be exploited to develop hydrogen storage devices. The work program in this manuscript has been divided into four sections. In the second section, we provide a brief explanation of the calculation scenario applied to this theoretical work. Details and observations concerning $A_2CrH_6$ (A = Ca, Sr, and Ba) are described in the third section. Within the final section, we summarize by highlighting the key points of this work.

## 2. Computational details

The Cambridge Serial Total Energy Package (CASTEP) program [15] was adopted as the simulation method to optimize the structure and examine the material properties of hydride materials $A_2CrH_6$ (A = Ba, Sr, and Ca). Using density functional theory (DFT) [19], the software CASTEP offers the ability to solve the Kohn-Sham equations [15], making it possible to accurately predict electronic and atomic properties. The exchange-correlation interactions were considered in the context of the generalized gradient approximation (GGA) [20], into which the Perdew-Burke-Ernzerhof (PBE) functional [21] was integrated, which has proven highly effective for comparable hydride structures. The ultra-soft pseudopotential plane wave procedure was chosen because of its speed of calculation and its compatibility with the reliable description of electron wave functions.

Periodic boundary conditions corresponding to Bloch's theorem [22] were taken into account to accurately model the crystal structures. To stimulate exchanges between electrons and ions while preserving an acceptable calculation cost, Vanderbilt ultrasoft pseudopotentials were selected. Structural optimization was implemented employing the Broyden-Fletcher-Goldfarb-Shanno (BFGS) minimization scheme [23], that iteratively reduces the total system energy of the system by matching both the wave functions and charge density to obtain the most stable configuration. Pulay's model was applied to enhance the convergence of the self-coherent field (SCF) [24] electronic cycle by rationally updating the charge density at all iterations, thereby improving convergence and accuracy. The Hubbard parameter U for chromium (Cr) atoms has a value of 4 eV [25]. For Brillouin zone modeling, a $12 \times 12 \times 12$ Monkhorst-Pack k-point grid was hired. An energy cutoff threshold ($E_{cut}$) of 600 eV for plane waves and an energy convergence threshold of $5 \times 10^{-6}$ eV/atom were defined in order to guarantee the validity of the structural and electronic responses. Structural stress relief was carried out until the maximum Hellmann-Feynman force on every atom was reduced to 0.01 eV/Å. To maintain structural integrity, the applied stress was constrained to 0.02 GPa during optimization. Furthermore, ionic displacement thresholds were set at $5 \times 10^{-4}$ Å to suppress unstable atomic motion and ensure convergence. The SCF and energy tolerances were adjusted to improve the accuracy of electronic structure computations. Following cell optimization, the structural and electronic characteristics of $Ca_2CrH_6$, $Sr_2CrH_6$, and $Ba_2CrH_6$ were systematically investigated.

## 3. Results and discussions

### 3.1. Structural stability

For our research, we looked at the hydride perovskites' structural features, namely their cubic symmetry structure, corresponding to the $Fm\bar{3}m$ space group (No. 225). As seen in **Figure 1**, these hydrides feature a cubic shape, in which the atoms grouped together inside the elementary cell at particular positions, as shown in **Table 1**.

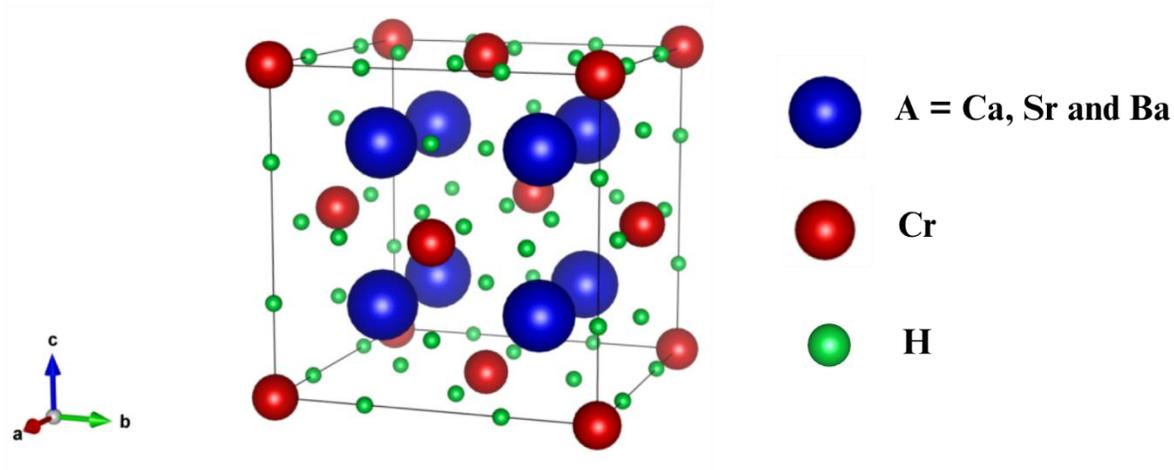

**Figure 1.** The perovskite elementary cell of A$_2$CrH$_6$ (A = Ca, Sr and Ba) perovskites.

**Table 1.** Atomic positions of hydrides A$_2$CrH$_6$ (A = Ca, Sr, and Ba) in the Fm-3m Space Group.

| Compound | | | Position | | |
|---|---|---|---|---|---|
| | Atom | Site | x | y | z |
| Ca$_2$CrH$_6$ | Ca | 8 c | 0.25 | 0.75 | 0.75 |
| | Cr | 4 a | 0 | 0 | 0 |
| | H | 24 e | 0.238330 | 0 | 0 |
| Sr$_2$CrH$_6$ | Sr | 8 c | 0.25 | 0.75 | 0.75 |
| | Cr | 4 a | 0 | 0 | 0 |
| | H | 24 e | 0.228190 | 0 | 0 |
| Ba$_2$CrH$_6$ | Ba | 8 c | 0.25 | 0.75 | 0.75 |
| | Cr | 4 a | 0 | 0 | 0 |
| | H | 24 e | 0.219640 | 0 | 0 |

Table 2 presents the important parameters obtained by structural relaxation, such as the optimized lattice parameter ($a_0$), unit cell volumes ($V_0$) and material densities ($\rho$). $Ba_2CrH_6$ has the highest lattice constant (8.082 Å) and unit cell volume (527.845 Å³), indicating barium's comparatively large atomic size in the atomic lattice [26]. In contrast, $Ca_2CrH_6$ presents the minimum lattice constant at 7.220 Å with an elementary cell volume of 376.303 Å³, denoting calcium's smaller atomic radius implies a denser atomic configuration. And let us not omit $Sr_2CrH_6$, which exhibits a cell parameter of 7.623 Å and a corresponding elementary cell volume of 442.952 Å³. Concerning density, $Ba_2CrH_6$ has the highest density (4.187 g/cm³), followed by $Sr_2CrH_6$, which has 3.498 g/cm³ and $Ca_2CrH_6$, which possesses a lower density of 2.439 g/cm³. This study demonstrates a comprehensible link among the metal atom's ionic radius and the crystallographic parameters: when the last-mentioned parameter increases, so does the cell volume and lattice parameter. Modifications to structural elements have an impact on stability, functionality, and hydrogen storage capacity in real-life uses [27].

Energy of formation ($\Delta E_f$) refers to the difference in energy required to produce a material from its constituent elements under optimal conditions. A negative value indicates that the molecule forms naturally and is thermodynamically stable [28]. The formation energies ($\Delta E_f$) for hydride materials $A_2CrH_6$ (A = Ca, Sr, and Ba) can be determined according to the equation below [9]:

$$\Delta E_f = \frac{E(A_2CrH_6) - (2E(A) + E(Cr) + 3E(H_2))}{9} \quad (1)$$

In the previous equation, $E(A_2CrH_6)$ refers to the total system energy of the material hydrides $A_2CrH_6$ (A = Ca, Sr, and Ba). Here, $E(A)$ denotes the total energy of the alkaline earth metal (A = Ba, Sr, and Ca), $E(Cr)$ indicates the total energy of the chromium atom, while $E(H_2)$ corresponds to the total energy of the hydrogen atom. **Table 2** shows the formation energies ($\Delta E_f$) for the hydride materials studied. The measured findings demonstrate that all these hydride compounds possess negative ($\Delta E_f$) values, suggesting their thermodynamic stability. $Ba_2CrH_6$ has the highest thermodynamic stability, with a formation energy of −0.631 eV/atom. All compounds' negative formation energies imply that the generation phase these hydrides is exothermic, implying system rejection into the environment [29].

**Table 2.** $A_2CrH_6$'s (A = Ca, Sr and Ba) structural properties, density and enthalpy of formation.

| Compound | $Ca_2CrH_6$ | $Sr_2CrH_6$ | $Ba_2CrH_6$ |
| --- | --- | --- | --- |
| **a** (Å) | 7.220 | 7.623 | 8.082 |
| **V** (Å³) | 376.303 | 442.952 | 527.845 |
| **ρ** (g/cm³) | 2.439 | 3.498 | 4.187 |
| **ΔE_f** (eV/atom) | −0.630 | −0.628 | −0.631 |

## 3.2. Phonon dispersion and thermal stability

To clarify the interactions of hydrogen with the host material in hydrogen storage hydrides, it is necessary to master the vibratory properties of hydrogen [30]. Optimizing hydrogen diffusion rates and storage capacity is based on these notions. The material's vibratory response provides an indication of its thermal stability. In order to define materials compatible with the temperature cycles associated with hydrogen absorption and desorption, it is important to determine its properties [31].

The dynamic stability of hydride perovskites $A_2CrH_6$ (A = Ca, Sr, and Ba) is assessed from phonon dispersion curves, because the oscillation frequencies of the atoms are also found on these curves. In this part, Phonon dispersion profiles and phonon density of states for these perovskite hydrides have been generated in this section using the CASTEP program [15] in conjunction with linear response theory and supercell methods. With 36 atoms in its cubic form, the hydride perovskites $A_2CrH_6$ (A = Ca, Sr, and Ba) has 108 phonon frequency curves. According to **Figure 2** (**a**, **c**, and **e**), the phonon frequencies for $A_2CrH_6$ (A = Ca, Sr, and Ba) range from 0 THz to 50 THz. Whereas the presence of imaginary frequencies would indicate dynamic instability, their absence often denotes dynamic stability. The dynamic stability of hydride materials $A_2CrH_6$ (A = Ca, Sr, and Ba) is highlighted in **Figure 2** [17], which verifies that their phonon spectra do not contain imaginary frequencies [17]. This finding is again confirmed by the phonon density of states (PhDOS), which is correlated with the phonon dispersion curves. **Figure 2** (**b**, **d** and **f**) displays the phonon density of states for $Ca_2CrH_6$, $Sr_2CrH_6$, and $Ba_2CrH_6$. This phononic spectrum clearly illustrates the distribution of atomic contributions to different frequency ranges in hydride materials $A_2CrH_6$ (A

= Ca, Sr, and Ba). The vertical axis represents the frequency (THz), which varies from 0 THz to 50 THz, while the horizontal axis indicates the density of states (states/eV). The curve shows the total density of states, while the partial density of states for atoms A (A = Ca, Sr, and Ba), Chromium (Cr) and Hydrogen (H). In the low-frequency region, vibrations are mainly governed by the heavier atoms, Ca and Cr, with only a small contribution from hydrogen. Lastly, hydrogen vibrations are also responsible for the peaks seen at high frequencies [32]. Thanks to their dynamic stability, the perovskite hydrides studied can be employed to develop hydrogen storage materials [33].

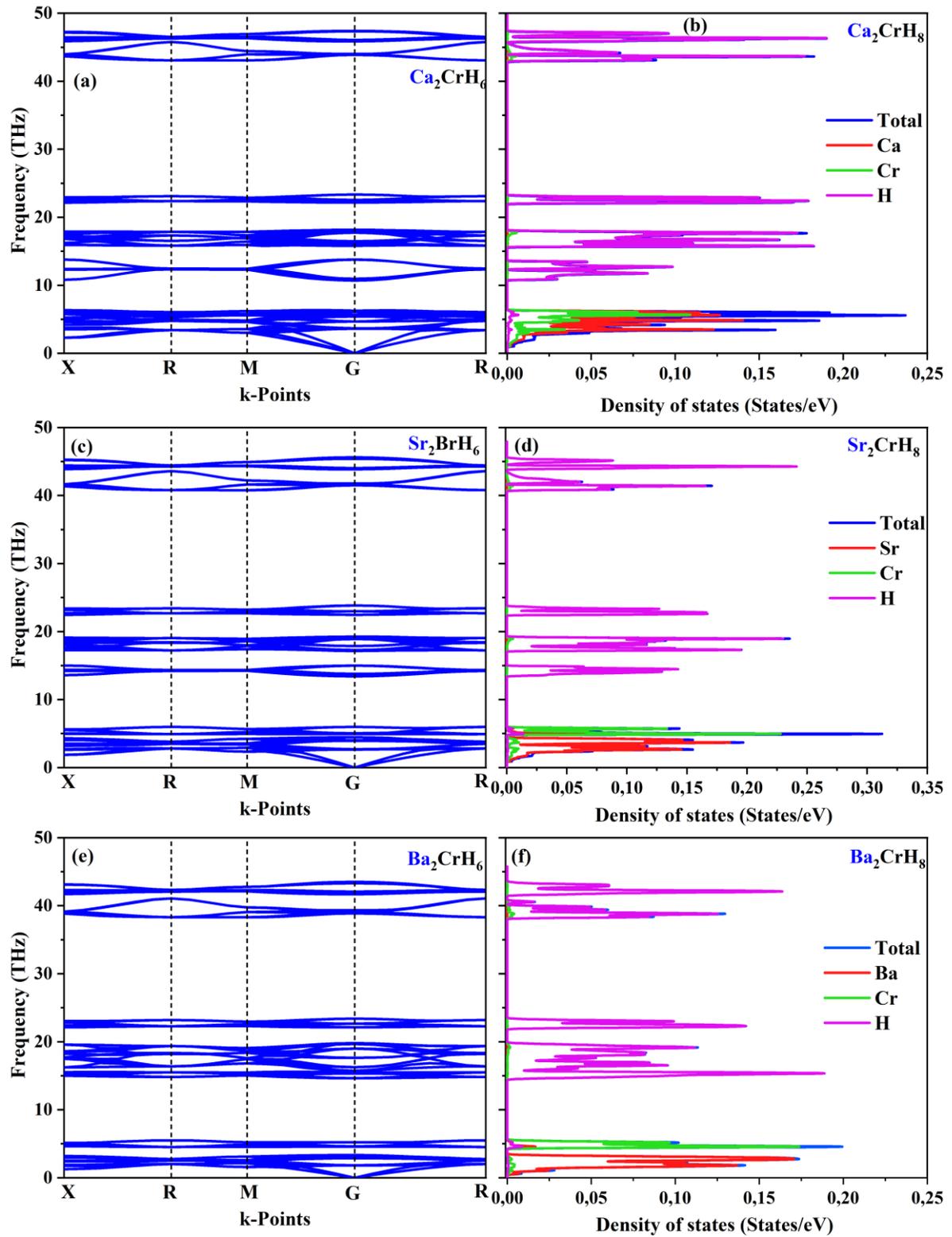

**Figure 2.** Vibrational behavior of $A_2CrH_6$ (A = Ca, Sr and Ba): A density functional theory approach.

Furthermore, we also present Ab-Initio Molecular Dynamics (AIMD) simulation results for these compounds to assess their thermal stability at 300 K. **Figure 3** illustrates AIMD simulation results for temperature fluctuations as a function of time in $Ca_2CrH_6$, $Sr_2CrH_6$, and $Ba_2CrH_6$. which provides essential information on the thermal properties of the three hydride compounds. The figures show recorded temperature variations with characteristic high-frequency oscillations around the mean value, throughout the simulations, the atomic mobility inside the studied material is represented directly. In the AIMD simulation, the temperatures of $A_2CrH_6$ (A = Ca, Sr, and Ba) change within defined limitations, demonstrating their intrinsic dynamic thermal reactivity.

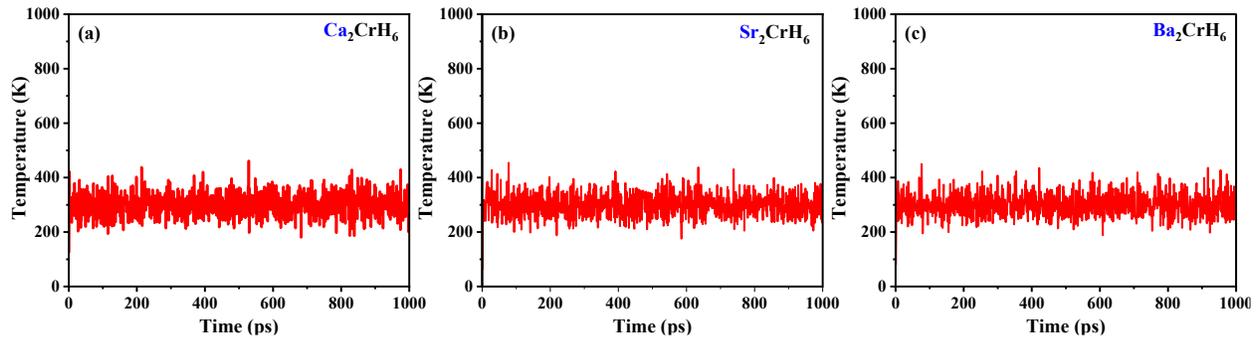

**Figure 3.** Temperature variations for perovskites $A_2CrH_6$ (A = Ca, Sr and Ba) depending on time.

**Figure 4** illustrates the energy evolution of hydrides $A_2CrH_6$ (A = Ca, Sr, and Ba) during an Ab Initio Molecular Dynamics (AIMD) [34] simulation at 300 K, indicating stable thermodynamic behavior [34]. Following a brief equilibration period (around 0 ps), the kinetic, potential, and total energies fluctuate around steady mean values without noticeable drift. This reflects efficient temperature control, rapid convergence, and stable thermodynamic conditions [35]. A clear anti-correlation between kinetic and potential energies is observed, characteristic of energy exchange within a thermostatted system. These results validate the simulation's reliability and support accurate analysis of the material properties. These findings show that hydride perovskites are thermally stable under temperature variations, which is essential for the study of their physical properties and likely applications in energy storage and perovskite research [36].

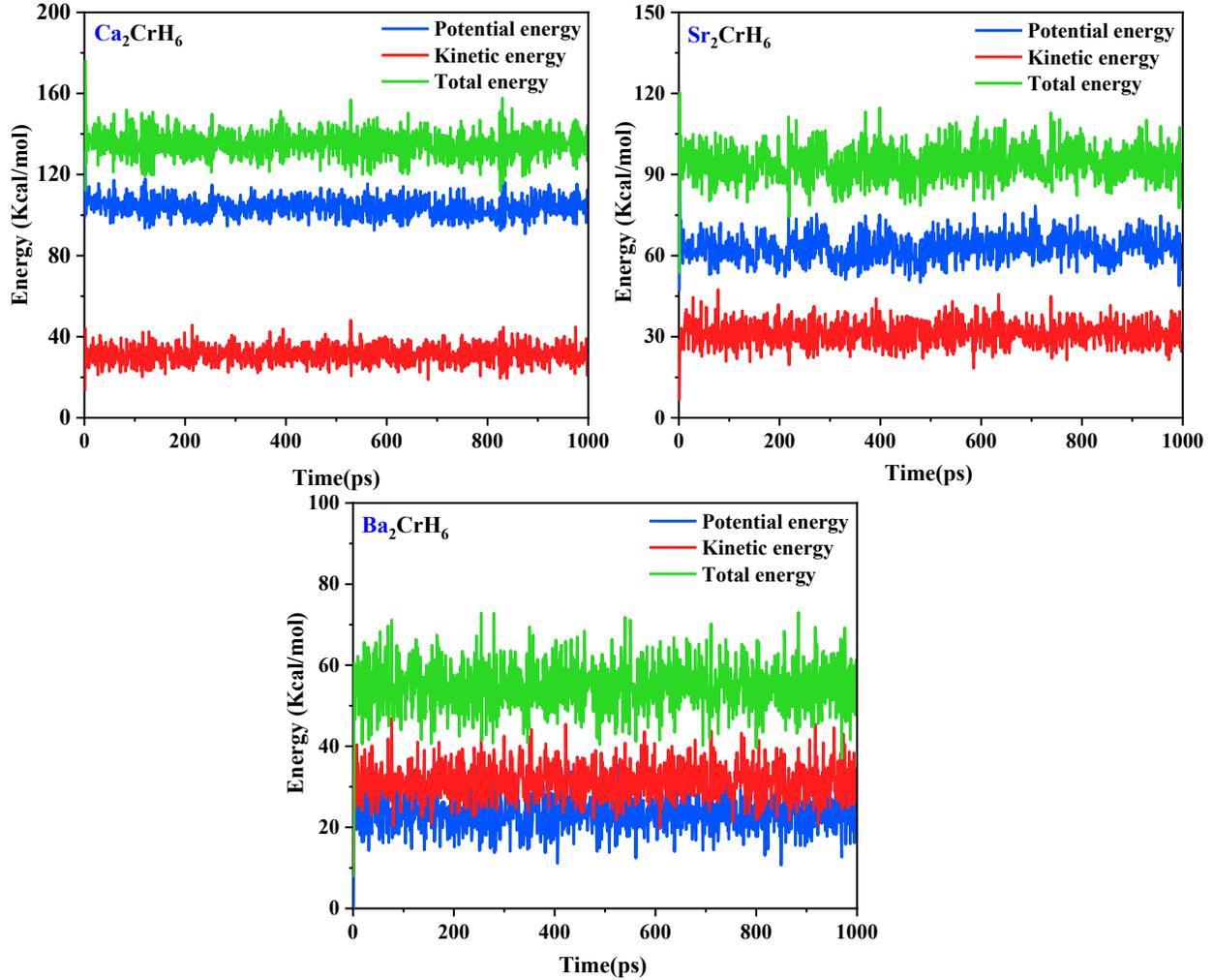

**Figure 4.** Total energy, Kinetic energy and Potential energy in relation to time in AIMD simulation at 300 K for perovskites $A_2CrH_6$ (A = Ca, Sr and Ba).

### 3.3. Thermodynamic properties

Applying phonon calculations within CASTEP's implementation of the quasi-harmonic approximation [15], the thermodynamic characteristics of hydride materials $A_2CrH_6$ (A = Ca, Sr, and Ba) were determined, suggesting their possible use for device design and production [37]. These hydrides' thermodynamic properties were calculated at a range of temperatures of 0 K - 1000 K. The objective of thermodynamic research is to study how different temperatures affect materials' properties. Various other physical properties are intrinsically related to the thermodynamic properties of materials, which can considerably influence the fundamental behavior of solid-state systems [38]. A deep knowledge of the materials' thermodynamic properties provides invaluable feedback on their performance under extreme conditions, such as high temperatures. Thermodynamic properties are crucial, and an optimal equilibrium between thermodynamic stability and modulation of the relevant parameters is required for efficient hydrogen storage.

The **Figure 5** plots the evolution of the thermodynamic properties (including enthalpy, Gibbs free energy and entropy) as a function of temperature for the hydride perovskites $A_2CrH_6$ (A = Ca, Sr, and Ba). The evolution of enthalpy, shown in **Figure 5 (a)**, indicates an approximately linear increase with temperature for all three compounds, the curves are very close to one another, suggesting a relatively low influence of the A (where A = Ca, Sr, and Ba) cation on this parameter. At 1000 K, enthalpy reaches 5.996 eV for $Ca_2CrH_6$, 6.010 eV for $Sr_2CrH_6$ and 6.106 eV for $Ba_2CrH_6$. In contrast, **Figure 5 (b)** clearly shows that the Gibbs free energy decreases significantly with increasing temperature [38], reflecting an improvement in thermodynamic stability at high temperatures [39]. Of the three materials, $Ba_2CrH_6$ exhibits the lowest free energy values over the entire range studied, indicating that it is the most thermodynamically stable, followed by $Sr_2CrH_6$ and $Ca_2CrH_6$. At 1000 K, the free-energy values are -7.687 eV for $Ca_2CrH_6$, -8.120 eV for $Sr_2CrH_6$ and -8.847 eV for $Ba_2CrH_6$ respectively, clearly showing that $Ba_2CrH_6$ is the most thermodynamically stable under these conditions. According to the laws of thermodynamics, **Figure 5 (c)** shows that the entropy effect, represented by the product T*entropy, increases with temperature. $Ba_2CrH_6$ has the highest values for this thermodynamic parameter, which then decreases in the $Sr_2CrH_6$ sequence, followed by $Ca_2CrH_6$. At 1000 K, this term is 13.386 eV for $Ca_2CrH_6$, 14.131 eV for $Sr_2CrH_6$ and 14.593 eV for $Ba_2CrH_6$. This can be explained by the increasing size of the A cation (where A = Ca, Sr, and Ba), which introduces more disorder and vibrational degrees of freedom into the crystal lattice. This joint analysis of the three thermodynamic quantities provides a deeper insight into the effect of the alkaline earth cation on the thermodynamic stability and entropic behavior of these compounds [40].

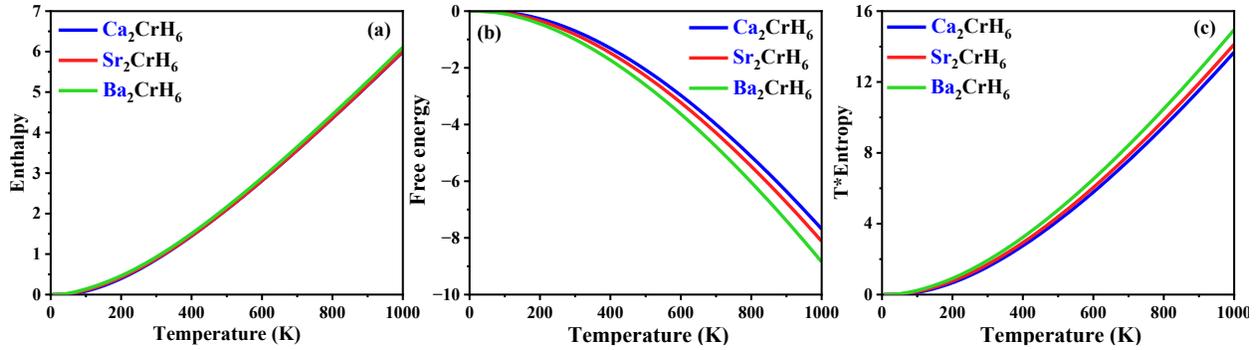

**Figure 5.** Graphical representation of (a) enthalpy, (b) free energy and (c) T*entropy of $A_2CrH_6$ (A = Ca, Sr and Ba).

Two significant parameters for comprehending the thermodynamic and vibratory behavior of solid materials are heat capacity and Debye temperature [41]. The quantity of the thermal energy necessary to elevate a material's temperature is indicated by its thermal capacity ($C_v$), while the maximum frequency of atomic vibrations and the rigidity of the crystal lattice are related to the Debye temperature ($\theta_D$) [30]. The **Figure 6** shows the evolution of these two properties (The thermal capacity and the Debye temperature) for hydride materials $A_2CrH_6$ (A = Ca, Sr, and Ba). Heat capacity increases rapidly at low temperatures, indicating the progressive activation of vibrational modes, then tends to stabilize at high temperatures, in accordance with Debye temperature. The three compounds' fairly similar trends show that their capacities to store thermal energy are equivalent. As for Debye temperature, the differences between all the compounds are more marked. At very low temperatures (5 K), $Ca_2CrH_6$ has the highest Debye temperature (368.315 K), reflecting high lattice rigidity, followed by $Ba_2CrH_6$ (314.028 K) and $Sr_2CrH_6$ (237.229 K). At high temperatures (1000 K), values increase significantly, to 1451.247 K, 1430.622 K and 1382.416 K for $Ca_2CrH_6$, $Sr_2CrH_6$ and $Ba_2CrH_6$, respectively. The ionic size of $A^{2+}$ (where A = Ca, Sr, and Ba) cations is reflected in this priority: a smaller cation (such as calcium ($Ca^{2+}$)) gives a more compact and rigid network, which increases the Debye temperature. Conversely, overall rigidity is reduced by the increased flexibility of the lattice due to the presence of a larger cation, such as barium ($Ba^{2+}$). These findings demonstrate that, despite similar thermal properties, lattice stiffness varies considerably depending on the type of alkaline earth cation [8]. This phenomenon can affect the dynamic, thermal, and mechanical properties of these hydrides in applications related to thermal conduction or energy storage [42].

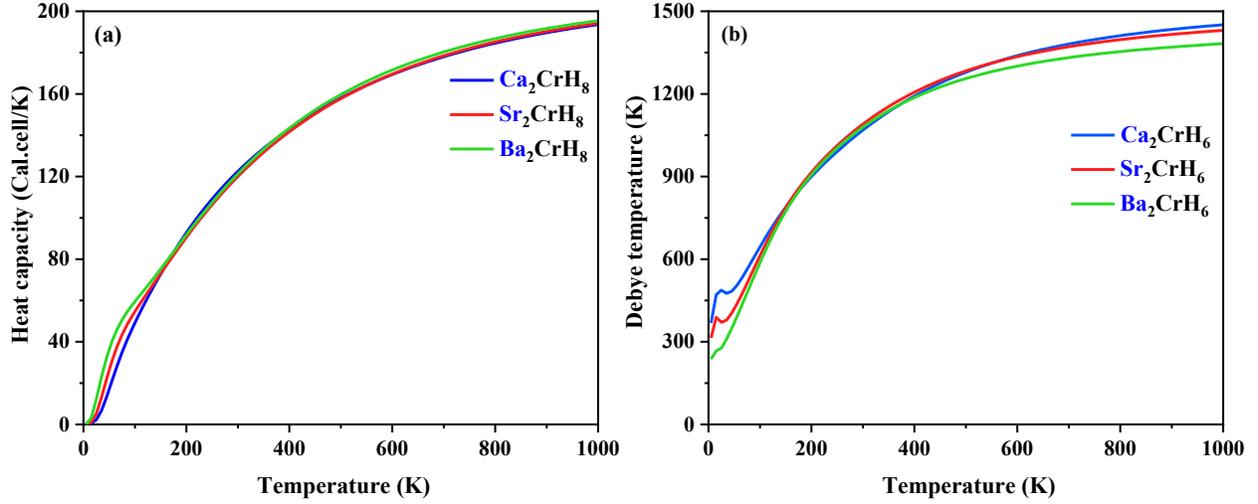

**Figure 6.** Heat capacity (a) and Debye temperature (b) of hydride perovskites $A_2CrH_6$ (A = Ca, Sr and Ba).

## 3.4. Elastic properties

A fundamental notion for understanding the mechanical behavior of materials, notably elastic anisotropy, fragility, bonding, ductility, resistance to fracture, stability and stiffness [43]. The mechanical constants of hydride compounds $A_2CrH_6$ (A = Ca, Sr, and Ba) can be determined with the CASTEP [15] with stress-strain technique, which offers a fuller idea of the mechanical properties of these materials. The hydride compounds $A_2CrH_6$ (A = Ca, Sr, and Ba), which have a crystal in the cubic system, have three elastic constants ($C_{11}$, $C_{12}$, and $C_{44}$). Based on Born's criterion, the mechanical stability of a compound can be accessed via the calculation criteria cited in the supplementary data [44].

$$C_{11} - C_{12} > 0 \quad ; \quad C_{11} + 2C_{12} > 0 \quad ; \quad C_{11} > 0 \quad ; \quad \text{and} \quad C_{44} > 0 \qquad (2)$$

In line with Born's criteria and the **Table 3**, the hydride perovskites $A_2CrH_6$ (A = Ca, Sr and Ba) are stable mechanically [45]. **Table 3** demonstrates the manner in which the alkaline earth cation elements present in the hydride $A_2CrH_6$ (A = Ca, Sr and Ba) affect their elastic constants.

Table 3. Elastic constants calculated for hydrides $A_2CrH_6$ (A = Ca, Sr and Ba).

| Materials | $C_{11}$ | $C_{12}$ | $C_{44}$ |
|---|---|---|---|
| $Ca_2CrH_6$ | 116.872 | 30.417 | 40.141 |
| $Sr_2CrH_6$ | 56.391 | 0.782 | 9.856 |
| $Ba_2CrH_6$ | 64.862 | 9.910 | 19.023 |

Table 4 compares the compounds' physical properties with those of previously reported hydrides. Mechanical properties including bulk modulus (B), shear modulus (G), Young's modulus (E), Poisson's ratio (v), Cauchy pressure ($C_p$), hardness (H) and Kleinman parameter (ζ) can be obtained from values derived from $C_{11}$, $C_{12}$ and $C_{44}$. The approximations of Voigt, Reuss and Hill have been adopted for these computations. Nonetheless, Hill's approximation is preferred for these mechanical modulus calculations, as it generates results that can be confirmed by experimental findings. Mechanical moduli were defined in the following expressions [17]:

$$B = \frac{C_{11}+C_{12}}{3} \quad (3)$$

$$G = \frac{G_V+G_R}{2} \quad (4)$$

$$G_V = \frac{C_{11}-C_{12}+3C_{44}}{5} \quad (5)$$

$$G_R = \frac{C_{11}-C_{12}+3C_{44}}{C_{44}+3(C_{11}-C_{12})} \quad (6)$$

$$E = \frac{9BG}{3B+G} \quad (7)$$

$$v = \frac{3B-2G}{2(3B+G)} \quad (8)$$

$$C_P = C_{11} - C_{12} \quad (9)$$

$$H = \frac{E(1-2v)}{6(1-v)} \quad (10)$$

$$\zeta = \frac{C_{11}+C_{12}}{7C_{11}-2C_{12}} \quad (11)$$

The parameters listed in **Table 4** provide valuable information on the mechanical properties of $A_2CrH_6$ (A = Ca, Sr, and Ba). The material's compressive resistance is termed its bulk modulus (B) [46], while the shear modulus (G) characterizes to the material's resistance to shearing [46]. Of the three compounds investigated, $Ca_2CrH_6$ has the highest bulk modulus (B = 59.236 GPa) and shear modulus (G = 41.348 GPa), suggesting a notably rigid structure and high resistance to shear stress. Conversely, $Sr_2CrH_6$ has the lowest bulk modulus (B = 19.318 GPa) and shear modulus (G = 15.161 GPa), reflecting considerable compressibility and an easily deformable structure. $Ba_2CrH_6$ exhibits moderate rigidity, with a bulk modulus of 28.227 GPa and an intermediate shear modulus of 22.049 GPa. This trend confirms that the introduction of heavier cations with larger ionic radii (Sr and Ba) reduces the shear strength of the crystal lattice.

Young's modulus (E), a key parameter describing the elastic response of a material, represents the stress to strain ratio in tension [45]. The highest possible value is observed for $Ca_2CrH_6$ (E = 100.630 GPa), which suggests good rigidity and minimal deformability. Inversely, the value for $Sr_2CrH_6$ is considerably lower (E = 36.052 GPa), while the lowest value is observed for $Ba_2CrH_6$ (E = 22.481 GPa), suggesting a significantly more flexible behavior. Evidently, the effect of the ionic radius increase on the elasticity of the material is demonstrated by this decrease in Young's modulus. A useful parameter for measuring the bonding properties of crystalline compounds is Poisson's ratio ($v$) [47]. Typical values spanning of 0.1 and 0.25 represent covalent and ionic bonds, respectively [47]. The measured Poisson's ratios for the hydride perovskite materials $A_2CrH_6$ (A = Ca, Sr, and Ba) ranging from 0.19 to 0.217 indicate a combination of covalent and ionic bonds. The hydride $Ca_2CrH_6$ ($v = 0.217$) tends towards more ionic behavior, while $Sr_2CrH_6$ ($v = 0.189$) and $Ba_2CrH_6$ ($v = 0.190$) are slightly more covalent.

**Table 4.** Mechanical properties and elastic moduli of hydride materials $A_2CrH_6$ (A = Ca, Sr and Ba).

| Parameters | $Ca_2CrH_6$ | $Sr_2CrH_6$ | $Ba_2CrH_6$ |
|---|---|---|---|
| B (GPa) | 59.236 | 19.318 | 28.227 |
| G (GPa) | 41.348 | 15.161 | 22.049 |
| E (GPa) | 100.630 | 36.052 | 22.481 |
| v | 0.217 | 0.189 | 0.190 |
| ζ | 0.405 | 0.158 | 0.301 |
| $C_p$ (GPa) | −9.722 | −9.074 | −9.114 |
| B/G | 1.433 | 1.274 | 1.280 |
| H | 8.525 | 4.787 | 6.207 |
| A | 0.929 | 0.354 | 0.692 |
| $v_m$ (m/s) | 4389.142 | 2487.527 | 2581.035 |
| θ (K) | 612.294 | 297.923 | 309.123 |
| $T_{melt}$ (K) | 1459.56 | 1145.76 | 1189.72 |

In addition, to determine whether materials are ductile or fragile, a factor named Cauchy pressure ($C_p = C_{12} - C_{44}$) is evaluated [48]. A positive $C_p$ value means it's ductile, whereas a negative value suggests it's brittle. For the perovskite materials $Ba_2CrH_6$, $Sr_2CrH_6$, and $Ca_2CrH_6$, the Cauchy modulus values $C_p$ retained are −9.114 GPa, −9.074 GPa, and −9.722 GPa, respectively. The fragility of these hydride perovskites is clearly highlighted by all these negative values, which may limit their application in sectors that demand a certain resistance to elastic deformation. The material's ductility is also evaluated by the Pugh ratio (B/G) [48]. According to this criterion, a value higher than 1.75 denotes ductile behavior, whereas a value below 1.75 signifies brittle behavior. The fragile behavior of $A_2CrH_6$ hydride perovskites (A = Ca, Sr, and Ba) is confirmed by the Pugh ratio (B/G).

One of the parameters used to evaluate material properties is the Kleinman parameter ($\zeta$) [10]. It characterizes the flexibility of a material under internal deformation and reflects the ability of atoms to displace under stress [10]. When Kleinman's parameter ($\zeta$) is higher, the mechanical properties and response to deformation are affected by easier internal atomic displacement. The Kleinman parameter values ($\zeta$) for $Ca_2CrH_6$, $Sr_2CrH_6$ and $Ba_2CrH_6$ are 0.405, 0.158 and 0.301, respectively. The internal flexibility of $Ca_2CrH_6$ is greater than that of $Sr_2CrH_6$ and $Ba_2CrH_6$, which may influence its ductility.

The following expressions establish additional factors such as anisotropy, Debye temperature, average wave velocity, and melting temperature for a better knowledge of the mechanic properties of hydride perovskites $A_2CrH_6$ (A = Ca, Sr, and Ba) [17]:

$$A = \frac{2C_{44}}{C_{11}-C_{12}} \qquad (12)$$

$$\theta_T = \frac{h}{k_B}\left(\frac{3n}{4\pi V_a}\right)^{\frac{1}{3}} \times v_m \qquad (13)$$

$$v_m = \left[\frac{1}{3}\left(\frac{1}{v_l^3} - \frac{1}{v_t^3}\right)\right]^{-\frac{1}{3}} \qquad (14)$$

$$v_l = \left(\frac{3B+4G}{3\rho}\right)^{\frac{1}{2}} \qquad (15)$$

$$v_t = \left(\frac{G}{\rho}\right)^{\frac{1}{2}} \qquad (16)$$

$$T_{melt} = [553 + 5.19C_{11}] \pm 300 \text{ K} \qquad (17)$$

A fundamental reference for the elastic anisotropy of the materials hydride studied $A_2CrH_6$ (A = Ca, Sr, and Ba) is the anisotropy factor (A). Estimated values of the anisotropy factor (A) correspond to variations in mechanical properties in different directions [49]. Mechanical properties change uniformly in all orientations when $A = 1$, meaning that the behavior is isotropic. When $A \neq 1$, mechanical properties change differently depending on orientation, which is reflected in anisotropic behavior [49]. The derived values for the anisotropy factor (A) are reported in **Table 4**. The values obtained for anisotropy factor (A) for hydride perovskites $A_2CrH_6$ (A = Ca, Sr, and Ba) are 0.929, 0.354 and 0.692 respectively. It can be observed that $Ca_2CrH_6$ exhibits quasi-isotropic behavior, its value being close to 1, while $Sr_2CrH_6$ shows marked anisotropy, and $Ba_2CrH_6$ moderately anisotropic behavior. This anisotropy has a potential impact on the scattering of elastic waves, mechanical deformation and directional stability of materials.

The average sound velocity ($v_m$) is an important indication of a material's rigidity: the higher the $v_m$, the stiffer the material [18]. For the hydride perovskite materials, the $v_m$ values are 4389.142 m/s for $Ca_2CrH_6$, 2487.527 m/s for $Sr_2CrH_6$, and 2581.035 m/s for $Ba_2CrH_6$. However, the rigidity of $Ca_2CrH_6$ is considerably better than that of $Sr_2CrH_6$ and $Ba_2CrH_6$, which have average velocities significantly inferior to those of $Ca_2CrH_6$. That suggests that the two hydrides $Ba_2CrH_6$ and $Sr_2CrH_6$ have more flexible structures, while the crystal lattice of $Ca_2CrH_6$ is correspondingly more rigid.

A significant scientific variable is the Debye temperature ($\theta_D$), which characterizes the oscillations of the crystal lattice and is associated with the thermal and mechanical properties of materials [50]. The Debye temperature values calculated for $Ba_2CrH_6$ are 2581.035 K, for $Sr_2CrH_6$ are 2487.527 K, and for $Ca_2CrH_6$ are 4389.142 K. It is clear in this case that the hydride $Ba_2CrH_6$ has a crystal lattice with much higher energy vibrations and greater rigidity compared to the other two. Debye's high temperature of $Ca_2CrH_6$ translates into potentially higher thermal conductivity and greater thermal stability.

The high melting point of the hydride is necessary to ensure the stability of the hydrogen stored under ideal conditions [50]. The melting temperature ($T_{melt}$) is 1459.56 K for $Ca_2CrH_6$, 1145.76 K for $Sr_2CrH_6$, and 1189.72 K for $Ba_2CrH_6$. The thermal stability of these perovskites is such that $Ca_2CrH_6$ can be melted at a temperature of $1273.15\ K$, which is a sign of improved resistance to thermal decomposition. In contrast, $Sr_2CrH_6$ and $Ba_2CrH_6$ materials attain lower melting temperatures, which means they have more limited thermal stability. These observations are in line with the mechanical properties observed, in specific the average velocity of sound, which suggest improved rigidity for hydride $Ca_2CrH_6$.

The ELATE software [51] was selected for its ability to execute a wide range of analyses, including determining material properties and calculating stress-strain graphs. The anisotropic elastic properties of three-dimensional hydride perovskites $A_2CrH_6$ (A = Ca, Sr, and Ba) are shown in **Figure 7**. This figure gives an idea of how the elastic modules are spread out, notably shear modulus (G), compressibility (B), Poisson's ratio (v), and Young's modulus (E). It provides an initial indication of the elastic anisotropy or isotropy of all materials, which is confirmed by the values shown in **Table 3.** The linear compressibility of the three materials has an anisotropy of 1, which corresponds to isotropic behavior, i.e., uniform behavior in all directions. This is particularly evident in three-dimensional images, where a spherical shape means that the minimum and maximum values are the same. In comparison, the other elastic properties show slight anisotropy.

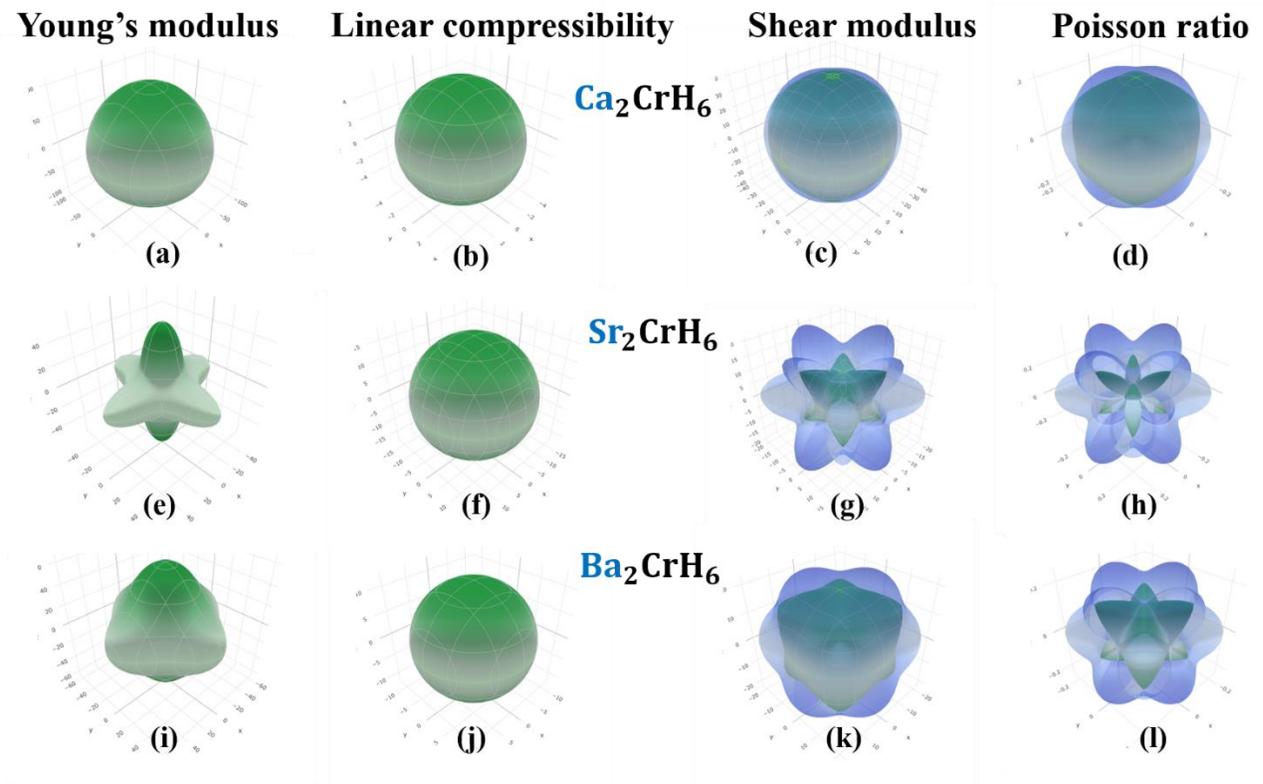

**Figure 7.** The anisotropic elastic properties of $Ca_2CrH_6$ (a, b, c, d), and $Sr_2CrH_6$ (e, f, g, h), and $Ba_2CrH_6$ (i, j, k, l) are shown in 3D.

**Table 5.** Young's modulus (E), compressibility (B), shear modulus (G), Poisson's ratio (v), and their anisotropy values for $A_2CrH_6$ (A = Ca, Sr and Ba) perovskites were calculated, along with their lowest and maximum values.

| Materials | E (GPa) | | | β (1/TPa) | | | G (GPa) | | | v | | |
|---|---|---|---|---|---|---|---|---|---|---|---|---|
| | $E_{min}$ | $E_{max}$ | Value Anisotropy | $β_{min}$ | $β_{max}$ | Value Anisotropy | $G_{min}$ | $G_{max}$ | Value Anisotropy | $v_{min}$ | $v_{max}$ | Value Anisotropy |
| $Ca_2CrH_6$ | 98.233 | 104.31 | 1.062 | 5.627 | 5.627 | 1.0 | 40.141 | 43.227 | 1.077 | 0.197 | 0.242 | 1.225 |
| $Sr_2CrH_6$ | 25.271 | 56.369 | 2.231 | 17.255 | 17.255 | 1.0 | 9.856 | 27.804 | 2.821 | 0.014 | 0.487 | 35.602 |
| $Ba_2CrH_6$ | 46.601 | 62.235 | 1.335 | 11.809 | 11.809 | 1.0 | 19.023 | 27.476 | 1.444 | 0.106 | 0.307 | 2.899 |

## 3.5. Electronic properties

To gain an insight into the electronic properties of a chosen hydride, it is important to study the band structure. This band structure explains how the electron energy values are allocated within a material and specifies the energy limits acceptable for electron occupation [52]. The study of electronic energy bands and energy gap enables us to forecast the thermal, electrical, and optical properties of a compound, which is necessary for the creation of materials designed for targeted applications [53]. **Figure 8** illustrates the band structures of hydride perovskites $A_2CrH_6$ (A = Ca, Sr, and Ba). This band structure were evaluated in the Brillouin zone employing the GGA-PBE method [20] [21], and the Fermi level is represented by a horizontal dotted path.

As observed in **Figure 8**, the spin-polarized electronic band structures of $A_2CrH_6$ (A = Ca, Sr, and Ba), separated into spin-up and spin-down channels, demonstrate pronounced electronic and magnetic activity [54]. The hydride material $Ca_2CrH_6$ exhibits a half-metallic behavior, with a semiconductor spin-up channel ($E_g$ = 2.005 eV) and a metallic spin-down channel, indicating conduction exclusively by one spin type. Similarly, the hydride $Sr_2CrH_6$ has a spin-up gap of 1.862 eV, with a metallic spin-down channel as well, confirming its half-metallic character [55]. In contrast, the perovskite material $Ba_2CrH_6$ is unique in that it exhibits metallic characteristics in both spin channels, making it a spin-polarized metal. The electronic profile depends on the alkaline earth cation (Ca, Sr, or Ba). This choice leads to a reduction in spin polarization and an upgrade in metallicity. Ba$_2$CrH$_6$ remains a strong hydrogen storage contender because of its metallic nature. This favors fast hydrogen diffusion inside the material. The metallic behavior also means excellent bonding, enabling the material to efficiently transfer heat and electricity. This is advantageous for rapid and efficient storage systems.

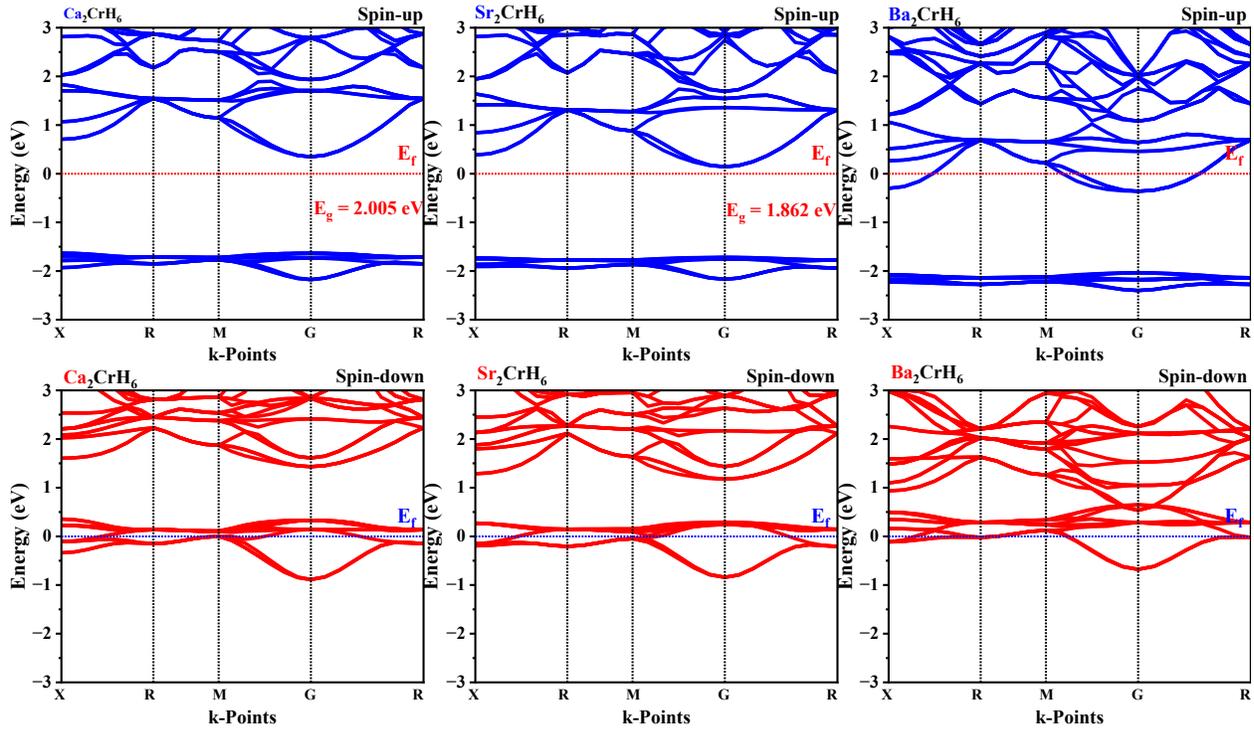

**Figure 8.** The electronic band structure of A$_2$CrH$_6$ (A = Ca, Sr and Ba) perovskites.

To fully explore the electronic behavior of a compound at the atomic level, we need to understand the total and partial density of states (TDOS and PDOS) [56]. This article highlights how individual orbitals such as s, p, d, and f impact the total electronic band structure of these compounds [57]. The TDOS and PDOS of the perovskite materials A$_2$CrH$_6$ (A = Ca, Sr, and Ba) are given in **Figure 9**. The Fermi level (E$_f$) represented by a black vertical line with dashes at 0 eV, which separates the valence band (VB) from the conduction band (CB).

The total and partial state densities of Ca$_2$CrH$_6$, Sr$_2$CrH$_6$ and Ba$_2$CrH$_6$ present remarkable structural similarities, even though their electronic and magnetic properties diverge. **Figure 3 (a, c**, and **e)** shows the total state density of hydride perovskites A$_2$CrH$_6$ (A = Ca, Ba, Mg, and Sr). Both compounds (Ca$_2$CrH$_6$ and Sr$_2$CrH$_6$) exhibit a visible band gap in the spin-up states, confirming their semiconducting nature for this spin state, while their spin-down channels exhibit states that cross the Fermi level, indicating metallic behavior. This asymmetry between the two spin states reflects strong spin polarization, indicating probable ferromagnetic magnetic order. In contrast, $Ba_2CrH_6$ exhibits metallic behavior in both spin channels: no band gap is observed in either spin-up or spin-down, reflecting the absence of spin polarization. Partial density of states (PDOS) reveals that Cr-d orbitals dominate the region close to the Fermi level, exerting a pivotal role in electronic and magnetic properties. The H-s orbitals contribute to the valence band, reflecting strong Cr-H hybridization. Alkaline-earth cations (Ca, Sr, and Ba), although not major contributors around the Fermi level, indirectly influence electronic properties via structural effects and hybridization modulation. Thus, $Ca_2CrH_6$ and $Sr_2CrH_6$ can be classified as magnetic semiconductors, while $Ba_2CrH_6$, metallic and non-polarized, is similar to a conventional metal. These properties, coupled with the presence of hydrogen and structural stability, render these hydrides compelling targets for hydrogen storage [58].

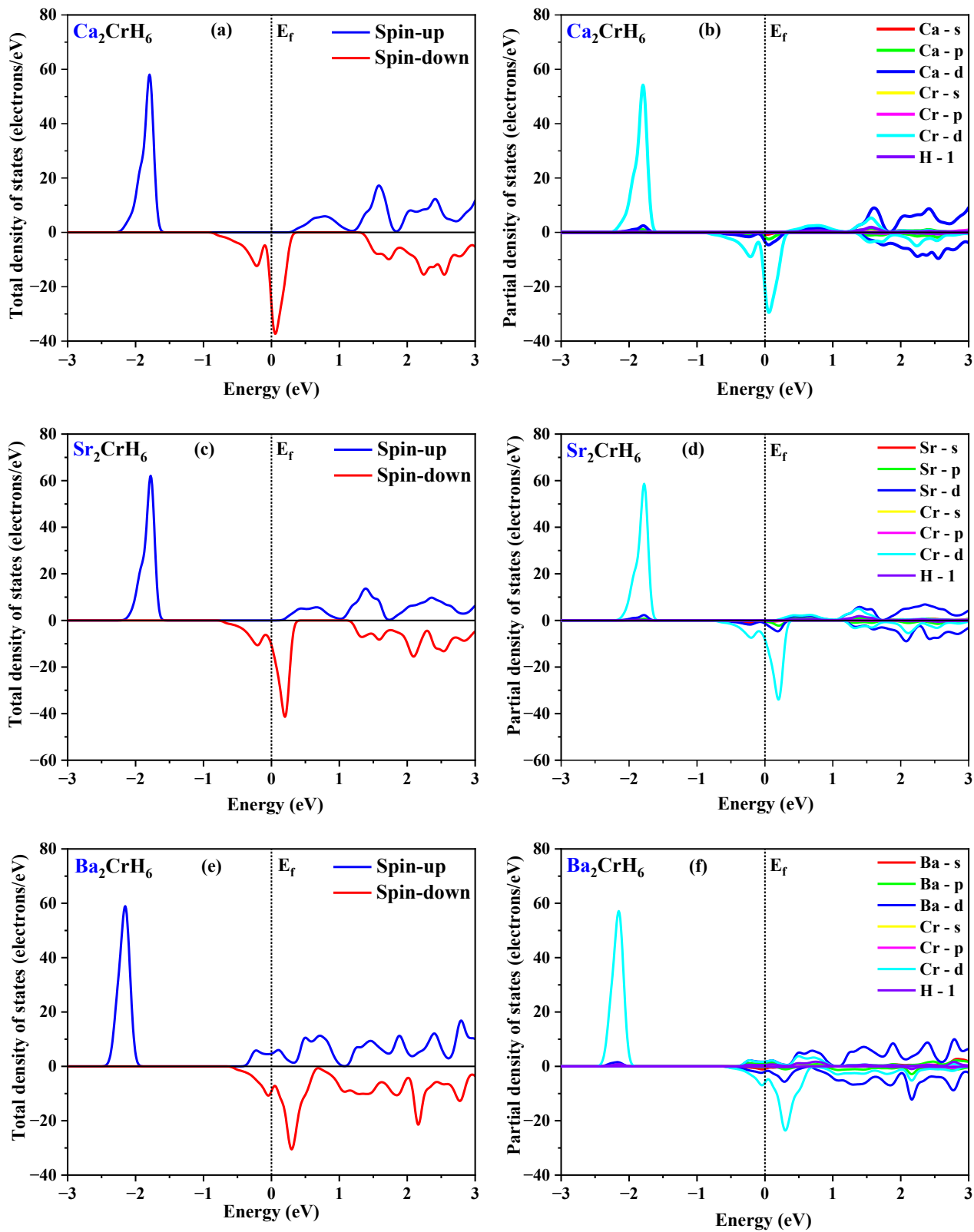

**Figure 9.** The computed DOS and PDOS of Ca$_2$CrH$_6$, Sr$_2$CrH$_6$ and Ba$_2$CrH$_6$ perovskites.

These properties, coupled with the presence of hydrogen and structural stability, render these hydrides an absolute choice for hydrogen storage. This finding is supported by Mulliken and Hirshfeld procedures [59], determining the magnetic moments and actual atomic charges of all atoms. These findings clarify the charge exchange between elements and the location of magnetism within the hydride structure. In particular, they confirm the observations made by studying the electronic bands and density of states [60]. They help to better comprehend the magnetic and electronic properties in these perovskite hydrides.

The results in **Table 6** indicate the magnetic moments and the atomic charges calculated by Mulliken and Hirshfeld methods [59] of the hydride materials $A_2CrH_6$ (A = Ca, Sr, and Ba). This complements the previous observations derived by band structures and DOS. The atomic charges represent the transition of charge from one element to another. Alkaline earth cations (Ba, Sr, and Ca) exhibit positive charges, supporting a role as electron givers. In addition, the hydrogen, functioning as a hydride ($H^-$), is negatively charged in all cases. Mulliken [59] discovered that chromium (Cr) gains electrons in $Ca_2CrH_6$ and $Sr_2CrH_6$, causing negative charges, but becomes a giver in $Ba_2CrH_6$, demonstrating a clear change in its electron exchange environment. The Hirshfeld method [59] assigns a positive charge to chromium (Cr) in all compounds, pointing to a more restrained electron reallocation and supporting its electropositive interaction with hydrogen. The magnetic moments indicate the magnetic nature of these materials: they are all non-zero, with values ranging from 2.22 $\mu_B$ ($Ba_2CrH_6$) to 2.59 $\mu_B$ ($Sr_2CrH_6$), indicating considerable spin polarization [25]. This polarization is consistent with the asymmetry found in the densities of states for up and down spins, meaning that magnetism is mainly carried by the chromium (Cr) atom. The difference in magnetic moments as a function of cation A is probably due to structural changes caused by increasing atomic size, which modifies local geometry, Cr-H hybridization and magnetic super exchange interactions. Remarkably, $Ba_2CrH_6$, despite being a metal in either spin state, retains a non-zero magnetic moment, making it a magnetic, but not spin-polarized, metal.

In summary, the atomic charges reveal an electronic redistribution consistent with the formation of complex hydrides, while magnetic moments demonstrate a significant interaction between electronic structure and magnetism. These properties, which can be controlled by the choice of A cation, increase the compounds' potential for use in spintronics and hydrogen storage [13].

**Table 6.** Electronic Structure Parameters of $A_2CrH_6$ (A = Ca, Sr and Ba): Charges and magnetic moment.

| Compound | Element | Charge(e) [Mulliken] | Charge(e) [Hirshfeld] | Magnetic moment |
|---|---|---|---|---|
| | Ca | 1.54 | 0.25 | |
| $Ca_2CrH_6$ | Cr | −1.57 | 0.19 | 2.307 |
| | H | −0.25 | −0.12 | |
| | Sr | 1.55 | 0.32 | |
| $Sr_2CrH_6$ | Cr | −1.46 | 0.17 | 2.587 |
| | H | −0.27 | −0.14 | |
| | Ba | 1.58 | 0.34 | |
| $Ba_2CrH_6$ | Cr | 0.27 | 0.19 | 2.220 |
| | H | −0.57 | −0.15 | |

### 3.6. Optical properties

Determining optical properties means understanding how materials interact with electromagnetic radiation [61], this is necessary to categorize materials and define potential uses, including certain hydrogen storage technologies, solar panels, optical layers and reflectors, and electronic devices. These light-matter interactions are explained by the dielectric function $\varepsilon(\omega)$ [62]. To identify the main optical properties, we calculated the real $\varepsilon_1(\omega)$ and imaginary $\varepsilon_2(\omega)$ components of the dielectric function from **Equation 18** [63].

$$\varepsilon(\omega) = \varepsilon_1(\omega) + \varepsilon_2(\omega) \tag{18}$$

The polarization of the material is expressed by the real component $\varepsilon_1(\omega)$, which demonstrates how the material delays and stores electromagnetic radiation without absorbing it [63]. Light absorption in the material is reflected by the imaginary fraction $\varepsilon_2(\omega)$, which corresponds to energy loss and dispersion during wave propagation [63].

$$\varepsilon_1(\omega) = 1 + \frac{2P}{\pi} \int_0^\infty \frac{\omega' \times \omega \times \varepsilon_2 \omega'}{\omega'^2 + \omega^2} d\omega \tag{19}$$

$$\varepsilon_2(\omega) = \frac{8}{2\pi\omega} \sum_{nn'} |P_{nn'}(k)| \frac{dS_k}{\nabla \omega_{nn'}(k)} \tag{20}$$

$$R(\omega) = \frac{n+ik-1}{n+ik+1} \tag{21}$$

$$I(\omega) = \sqrt{2\omega} \left[ \sqrt{\varepsilon_1^2(\omega) + \varepsilon_2^2(\omega)} - \varepsilon_1(\omega) \right]^{\frac{1}{2}} \tag{22}$$

$$n(\omega) = \frac{1}{\sqrt{2}} \left[ \frac{\sqrt{\varepsilon_1^2(\omega) + \varepsilon_2^2(\omega)} + \varepsilon_1(\omega)}{2} \right]^{\frac{1}{2}} \tag{23}$$

$$K(\omega) = \frac{1}{\sqrt{2}} \left[ \sqrt{\varepsilon_1^2(\omega) + \varepsilon_2^2(\omega)} - \varepsilon_1(\omega) \right]^{\frac{1}{2}} \tag{24}$$

$$\delta(\omega) = \frac{\alpha(\omega) n(\omega) c}{4\pi} \tag{25}$$

$$L(\omega) = \mathrm{Im}(\varepsilon(\omega)^{-1}) = \frac{\varepsilon_2^2(\omega)}{\varepsilon_1^2(\omega)} + \varepsilon_2^2(\omega) \tag{26}$$

According to **Equation 18** [63], the dielectric function $\varepsilon(\omega)$ consists of two primary components: the real part $\varepsilon_1(\omega)$ and the imaginary part $\varepsilon_2(\omega)$. The variation of the real part $\varepsilon_1(\omega)$, which represents the material's response to electromagnetic radiation, is shown in **Figure 10 (a)**. Specifically, it reflects how photons propagate through the material and how easily the electron cloud can be polarized when exposed to an electric field. The energy spectrum of $\varepsilon_1(\omega)$, ranging from 0 eV to 12 eV, offers valuable insights into the optical properties of the material. This graph demonstrates the evolution of the real part of the dielectric function $\varepsilon_1(\omega)$, as a function of energy for the perovskites $A_2CrH_6$ (A = Ca, Sr and Ba). At low energies, $\varepsilon_1(\omega)$ attains high values for all three compounds, indicating marked polarizability. The static values of the imaginary part of the dielectric function $\varepsilon_1(0)$ are are high, with values of 61.288 for $Ca_2CrH_6$, 32.578 for $Sr_2CrH_6$ and 25.838 for $Ba_2CrH_6$. In the visible range, $\varepsilon_1(\omega)$ remains positive and greater than 1, confirming the transparency of the compounds in this spectral range. A peak is observed at higher energies (in the ultraviolet) around 4 eV and 5 eV, linked to interband electronic transitions that signal the onset of optical absorption. The three hydrides exhibited comparable optical behaviors, with slight variation due to the difference in size of the alkaline-earth cations, subtly influencing their electronic polarization and interaction with light. **Figure 10 (b)** shows the imaginary part of the dielectric function $\varepsilon_2(\omega)$, of $A_2CrH_6$ (A = Ca, Sr and Ba) hydrides as a function of energy. The static values of the imaginary part of the dielectric function $\varepsilon_2(0)$ are 20.747 for $Ca_2CrH_6$, 9.516 for $Sr_2CrH_6$ and 4.471 for $Ba_2CrH_6$, indicating a higher polarizability for $Ca_2CrH_6$, reflecting a stronger interaction with a static electric field [64]. At low energies, there is significant absorption, especially pronounced for $Ca_2CrH_6$, related to infrared phenomena. In the visible region, $\varepsilon_2(\omega)$ remains very low and approaching zero for all three hydrides, affirming their transparency in this spectral range. In the ultraviolet range, absorption rises markedly, with several peaks associated with interband electronic transitions. The principal peak, between 5.5 eV and 6.5 eV, is strongest for $Ca_2CrH_6$, suggesting stronger UV absorption for compounds based on lighter elements.

The absorption coefficient, derived from the real and imaginary parts of the dielectric function (**Equation 18** [63]), indicates the efficiency with which a material absorbs photons and favors electronic transitions from the valence band to the conduction band [65]. **Figure 10 (c)** shows the optical absorption coefficient of hydrides $A_2CrH_6$ (A = Ca, Sr and Ba) as a function of energy. In the visible range, absorption coefficients are low, confirming the transparency of all three perovskite materials. However, in the ultraviolet, the absorption coefficient increases sharply: $Sr_2CrH_6$ has the most intense peak (206183.94 cm$^{-1}$ at 8.299 eV), followed by $Ca_2CrH_6$ (191293.7 cm$^{-1}$ at 7.988 eV), while $Ba_2CrH_6$ reaches a less pronounced maximum of 161729.301 cm$^{-1}$ at 7.528 eV. In practical terms, $Sr_2CrH_6$ would be the best of the three for applications such as UV filtering, photodetectors or UV-driven photocatalysis, which require high UV light absorption.

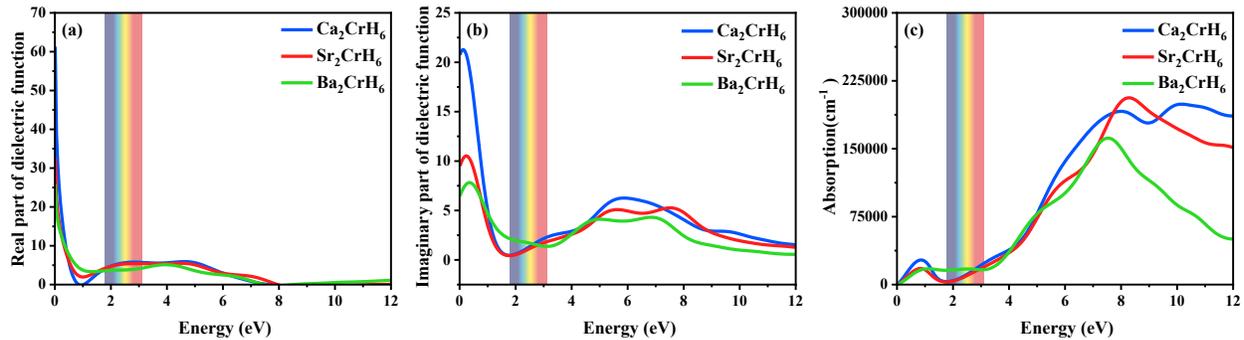

**Figure 10.** The real part (a) and imaginary part (b) of dielectric function and absorption coefficient (c) of $A_2CrH_6$ (A = Ca, Sr and Ba) perovskites.

**Figure 11 (a)** shows the refractive index (n) of perovskite hydrides $A_2CrH_6$ (A = Ca, Sr and Ba) as a function of energy. The static refractive index values are 7.937 for $Ca_2CrH_6$, 5.767 for $Sr_2CrH_6$ and 5.122 for $Ba_2CrH_6$ respectively. In the visible region, the index is greater than 1, with values between 2 and 2.5, indicating good transparency and strong light slowing [66], particularly for $Ca_2CrH_6$ and $Sr_2CrH_6$. However, in the ultraviolet, the refractive index decreases due to the dispersion associated with absorption peaks, reflected in the imaginary dielectric function. The behavior of $\varepsilon_2(\omega)$ and the extinction coefficient $k(\omega)$ observed in **Figure 11 (b)** are similar. The extinction coefficient $k(\omega)$ of hydride perovskites $A_2CrH_6$ (A = Ca, Sr and Ba) shows a main peak in the infrared: from 2.405 to 0.482 eV for $Ca_2CrH_6$, 1.576 to 0.522 eV for $Sr_2CrH_6$ and from 1.206 to 0.610 eV for $Ba_2CrH_6$. In the visible region, the extinction coefficient is less than 0.6 for all three materials, indicating good transparency and order of clarity [66]. In contrast, in the ultraviolet, all exhibit marked absorption, linked to interband electronic transitions, making them potential UV filters, with a bandgap probably lower for $Ba_2CrH_6$.

**Figure 11 (c)** shows the reflectivity of materials $A_2CrH_6$ (A = Ca, Sr and Ba) as a function of energy. In the infrared range, reflectivity is high, with static values of 0.611 for $Ca_2CrH_6$, 0.504 for $Sr_2CrH_6$ and 0.459 for $Ba_2CrH_6$; it then decreases sharply as it approaches the visible, suggesting potential for IR applications [67]. In the visible range, reflectivity is low than 0.2 for all three materials, indicating that they absorb more light than they reflect, giving them an unmetallic appearance; $Ba_2CrH_6$ has the lowest value. In the ultraviolet zone, reflectivity increases again, reaching peaks of 0.300 around 7.516 eV for $Ca_2CrH_6$, 0.317 around 8.299 eV for $Sr_2CrH_6$, and 0.262 around 7.666 eV for $Ba_2CrH_6$. $Sr_2CrH_6$ shows the highest reflectivity in the far UV, while $Ba_2CrH_6$ remains lower.

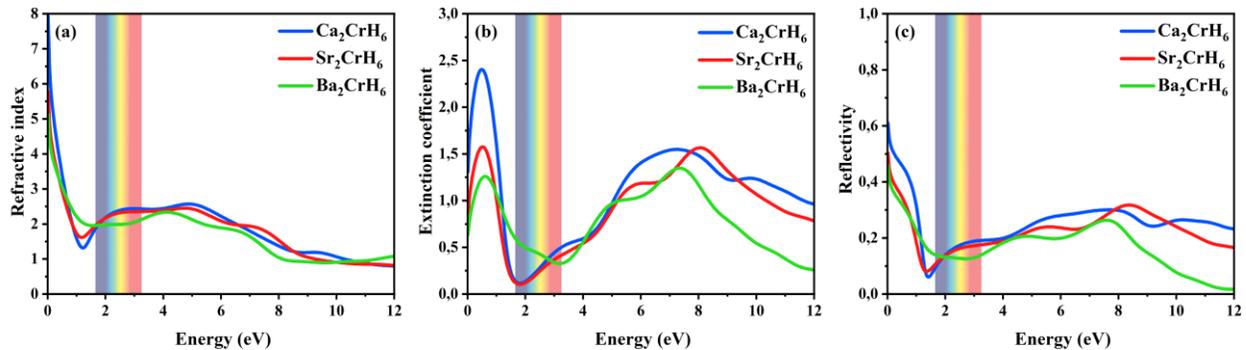

**Figure 11.** The refractive index (a), Extinction coefficient (b) and reflectivity (c) of $A_2CrH_6$ (A = Ca, Sr and Ba) perovskites.

**Figure 12 (a)** presents the real part of the optical conductivity of perovskite hydrides $A_2CrH_6$ (A = Ca, Sr and Ba) as a function of energy (eV). At low energies, non-zero conductivity is observed with small peaks around 0.576 eV ($Ca_2CrH_6$) and 0.662 eV ($Sr_2CrH_6$), possibly due to low-energy electronic transitions. In the visible range, optical conductivity is low (less than 1) for all three materials, indicating low absorption and therefore transparency or semi-transparency; $Ba_2CrH_6$ has a slightly higher conductivity than the other two, suggesting slightly lower transparency. In the ultraviolet, conductivity increases sharply, with peaks from 4.729 to 6.761 eV for $Ca_2CrH_6$, from 4.482 to 7.693 eV for $Sr_2CrH_6$, from 3.615 to 7.020 eV for $Ba_2CrH_6$, corresponding to interband electronic transitions; $Sr_2CrH_6$ shows the most intense peak, while $Ba_2CrH_6$ shows the weakest peak, reflecting differences in electronic structure. All things considered, these materials are weak absorbers in the visible spectrum, but powerful absorbers in the UV spectrum [68].

**Figure 12 (b)** illustrates the imaginary part of the optical conductivity $\sigma_2(\omega)$ of hydride compounds $A_2CrH_6$ (A = Ca, Sr and Ba). At low energies, imaginary conductivity is low and approaching zero, reflecting stable optical behavior [68]. In the visible, $\sigma_2(\omega)$ is negative for all three hydrides, indicating a dielectric behavior where polarization lags behind the electric field [69]. In the ultraviolet, $\sigma_2(\omega)$ follows its decrease to values: -2.830, -2.530 and -2.037 for $Ca_2CrH_6$, $Sr_2CrH_6$ and $Ba_2CrH_6$, respectively. Then it becomes positive at energies 6.950 eV, 7.600 eV and 6.974 eV for $Ca_2CrH_6$, $Sr_2CrH_6$ and $Ba_2CrH_6$, respectively. $\sigma_2(\omega)$ exhibits peaks from 1.393 at 8.413 eV for $Ca_2CrH_6$, from 1.813 at 8.671 eV for $Sr_2CrH_6$, and from 1.234 at 7.943 eV for $Ba_2CrH_6$, reflecting strong energy absorption consistent with the real part of the conductivity and the extinction coefficient [70].

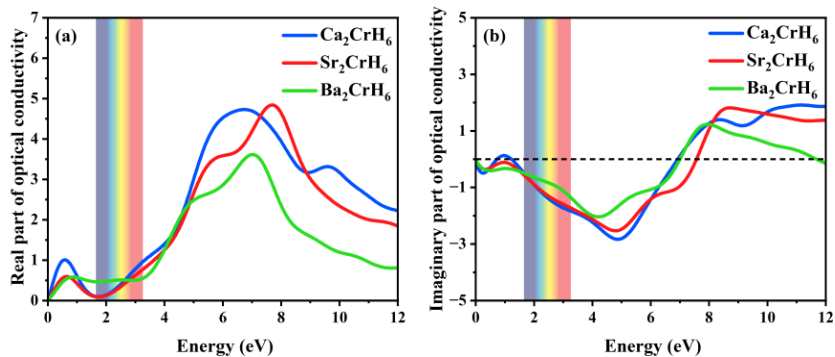

**Figure 12.** The real part (a) and imaginary part (b) of optical conductivity of $A_2CrH_6$ (A = Ca, Sr and Ba) perovskites.

## 3.7. Hydrogen storage

Storing hydrogen in a reliable form at a reasonable price is considered a technological obstacle to its integration on a large scale as a sustainable energy source [45]. There are several classes of hydrogen storage hydrides, and these classes seem to be reshuffled regularly as new hydride materials are produced that contradict the old classes [9]. To overcome the constraints imposed by standard techniques for storing hydrogen in compressed or liquid form, new ideas using advanced materials such as metal hydrides have been introduced. However, these devices cannot effectively control adsorption and desorption, nor can they deliver effective hydrogen storage capacity. Hydrogen has the greatest calorific value per mass of all synthetic chemical fuels. High-performance hydrogen storage is fundamental to the development of hydrogen energy markets, involving distribution systems and initial markets [71]. Among the various hydrogen storage technologies, adsorption on carbon nanotubes and solid phase technology stand out for their efficiency in terms of storage density, making them strategic options for the future development of the hydrogen economy [72]. **Table 7** summarizes the gravimetric and volumetric storage capacities of perovskite hydrides $A_2CrH_6$ (A = Ca, Sr, Ba). Gravimetric storage capacity, defined as the ability of a compounds to absorb hydrogen per unit weight, is defined as follows [47]:

$$C_{wt.\%} = \left(\frac{mM_H}{M_{host}+mM_H} \times 100\right)\% \qquad (18)$$

In this case, m denotes the proportion of hydrogen atoms in the host material, while $M_H$ and $M_{Host}$ correspond to the molar masses of the hydrogen atom and the host material, respectively [48]. The gravimetric storage capacity ($C_{wt.\%}$) in the compounds $Ca_2CrH_6$, $Sr_2CrH_6$ and $Ba_2CrH_6$ are 4.37 wt.%, 2.69 wt.% and 1.82 wt.%, respectively, demonstrating a gradual decrease in gravimetric storage capacity from Calcium (Ca) to Barium (Ba). Variations in the compounds' specific surface area, crystal structure, or chemical reactivity might be the cause of this tendency. The hydride $Ca_2CrH_6$ can be characterized by high sensitivity to the synthesis environment.

Additionally, we have calculated the volumetric storage capacity of hydrogen. The quantity of hydrogen (usually expressed in grams or liters) that a material can store per unit volume is known as its volumetric storage capacity [73]. It evaluates a material's potential to store hydrogen in a specific space, which is crucial for sectors such as hydrogen storage, where space is at a premium.

The volumetric hydrogen storage capacities of the perovskite hydrides $A_2CrH_6$ (where A = Ca, Sr, and Ba) are compiled in **Table 7**. Hydride perovskites $A_2CrH_6$ (A = Ca, Sr, and Ba) exhibit volumetric hydrogen storage capacities of 106.6 $gH_2/L$, 90.6 $gH_2/L$ and 76.2 $gH_2/L$, respectively. From these results, $Ca_2CrH_6$ is the most promising of the three candidates for applications needing efficient, compact hydrogen storage, as it has the highest density of hydrogen stored per unit volume. Conversely, $Ba_2CrH_6$ offered the lowest capacity, which may limit its interest in devices where storage density is a decisive criterion [74].

The temperature at which an atom that has been adsorbed to a material's surface or absorbed into its structure is released is known as the desorption temperature [75]. It is absolutely necessary to identify this temperature accurately when considering the performance of a hydrogen storage material and its suitability for various uses. The following formula is valid for deriving the hydrogen desorption temperature of hydride materials $A_2CrH_6$ (A = Ca, Sr, and Ba) [16]:

$$T_{des} = \frac{\Delta E_f}{\Delta S} \tag{19}$$

$\Delta E_f$ and $\Delta S$ stand for the formation energy and entropy changes, respectively, in the equation above, the entropy change of hydrogen materials is generally set at approximately -130.7 J/mol [16]. **Table 7** summarizes the hydrogen desorption temperatures of perovskite hydrides $A_2CrH_6$ (A = Ca, Sr, Ba). Hydrogen desorption temperatures for hydride materials $A_2CrH_6$ (A = Ca, Sr, and Ba) are given here: $Ca_2CrH_6$ at 465.2K, $Sr_2CrH_6$ at 463.7K, and $Ba_2CrH_6$ at 466.1K, indicate that the compound liberates hydrogen at comparatively elevated temperatures. Such extreme temperatures make these materials impractical for practical use, since the US Department Of Energy (DOE) [76] recommends desorption temperatures below 358 K for portable or embedded systems. As a result, the effectiveness and efficiency of this hydride in hydrogen storage systems is probably limited by the high energy it requires for desorption [77].

**Table 7.** $A_2CrH_6$ (A = Ca, Sr and Ba) gravimetric storage capacity and desorption temperature.

| Compound | $Ca_2CrH_6$ | $Sr_2CrH_6$ | $Ba_2CrH_6$ |
|---|---|---|---|
| $C_{wt.\%}$ | 4.37 | 2.69 | 1.82 |
| $\rho$ ($gH_2/L$) | 106.6 | 90.6 | 76.2 |
| $T_{des}$ (K) | 465.2 | 463.7 | 466.1 |

## 4. Conclusion

Molecular dynamics simulations and density functional theory (DFT) calculations implemented in the CASTEP program have permitted an extensive study of the structural, hydrogen storage, mechanical, electronic, and optical properties of hydride perovskites $A_2CrH_6$ (A = Ca, Sr, and Ba). The lattice constants of these hydride materials are 7.220 Å for $Ca_2CrH_6$, 7.623 Å for $Sr_2CrH_6$, and 8.082 Å for $Ba_2CrH_6$, indicating that they have stable cubic crystal structures. Phonon dispersion, negative formation energies, AIMD simulations, and elastic constants testify to their dynamic, thermodynamic, thermal, and mechanical stability, respectively. With an attractive desorption temperature of 465.2 K and a remarkable gravimetric hydrogen storage capacity of 4.37% by weight, $Ca_2CrH_6$ is a suitable choice for hydrogen storage applications. The electronic bands show remarkable spin activity, demonstrating that the change of $A+$ cation (where A = Ca, Sr, and Ba) immediately influences the spin polarization and electronic behavior of hydride perovskites. The hydrides $Sr_2CrH_6$ and $Ba_2CrH_6$ exhibit a slightly more covalent bond type, while $Ca_2CrH_6$ has a relatively ionic bond, according to Poisson's ratio. The ability to adapt to optoelectronic devices is indicated by certain optical parameters, such as refractive index, dielectric function, and absorption. Overall, these studies show that hydride perovskites $A_2CrH_6$ (A = Ca, Sr, and Ba), and in particular $Ca_2CrH_6$, are viable choices for advanced energy applications and hydrogen storage systems due to their high stability and adaptable electronic and optical properties.

### REFERENCES

[1] A. M. Ali et al., "A DFT investigation of Li-based hydride perovskites $LiXH_3$ (X = S, Se) for hydrogen storage applications," Int. J. Hydrog. Energy, vol. 154, p. 150299, Aug. 2025, doi: 10.1016/j.ijhydene.2025.150299.

[2] K. Abe, L. A. Lesmana, and M. Aziz, "Exploring the potential of high-performance hydrogen storage using metal hydrides with triply periodic minimal surface structures for mobility applications," Int. J. Hydrog. Energy, vol. 166, p. 150831, Sept. 2025, doi: 10.1016/j.ijhydene.2025.150831.

[3] R. Oualaid et al., "Study of mechanical, optical, electrical and structural properties of magnesium-based double perovskites $Mg_2XH_6$ (X= V, Cr) for hydrogen storage applications using DFT," Solid State Commun., vol. 404, p. 116102, Oct. 2025, doi: 10.1016/j.ssc.2025.116102.

[4] S. K. Yadav, U. Chaudhary, and G. C. Kaphle, "Exploring the structural, elastic, electronic, and optical properties of Na-based hydrides $X_3NaH_4$ (X = K, Rb) for hydrogen storage applications: A First-principles study," Int. J. Hydrog. Energy, vol. 160, p. 150637, Aug. 2025, doi: 10.1016/j.ijhydene.2025.150637.

[5] S. Zhang et al., "First principles study on the structure, hydrogen storage, and physical properties of $X_2VH_6$ (X = Mg, Ca, Sr, Ba) perovskite hydrides for hydrogen storage applications," Int. J. Hydrog. Energy, vol. 166, p. 150903, Sept. 2025, doi: 10.1016/j.ijhydene.2025.150903.

[6] G. M. Mustafa et al., "First principle study of physical aspects and hydrogen storage capacity of magnesium-based double perovskite hydrides $Mg_2XH_6$ (X= Cr, Mn)," Int. J. Hydrog. Energy, vol. 95, pp. 300–308, 2024.

[7] Z. Duan et al., "Reversible hydrogen storage with Na-modified Irida-Graphene: A density functional theory study," Int. J. Hydrog. Energy, vol. 85, pp. 1–11, 2024.

[8] Q. Dai, T.-Y. Tang, Q.-Q. Liang, Z.-Q. Chen, Y. Wang, and Y.-L. Tang, "Exploration of $A_2BH_6$ (A = K, Rb; B = Ge, Sn) hydrides for hydrogen storage applications: A first principles study," Int. J. Hydrog. Energy, vol. 92, pp. 769–778, Nov. 2024, doi: 10.1016/j.ijhydene.2024.10.324.

[9] R. Zosiamliana et al., "A comprehensive first principles investigation of $A_2BH_6$ type (A= Li,Na, and K; B= Al, and Si) double perovskite hydrides for high capacity hydrogen storage," July 26, 2025, arXiv: arXiv:2507.19810. doi: 10.48550/arXiv.2507.19810.

[10] K. Sabir, K. EL-Achouri, L. Bahmad, and O. E. Fatni, "Investigation of the structural, electronic, mechanical, and optical properties of $X_2MgH_6$ (X=Ba, Sr) hydrides for hydrogen storage," Intellect. J. Energy Harvest. Storage, vol. 3, no. 1, pp. 55–68, June 2025, doi: 10.11591/ehs.v3i1.pp55-68.

[11] A. Almahmoud, H. Alkhalidi, and A. Obeidat, "Comprehensive DFT analysis of structural, mechanical, electronic, optical, and hydrogen storage properties of novel perovskite-type hydrides $Y_2CoH_6$ (YCa, Ba, Mg, Sr)," J. Energy Storage, vol. 117, p. 116146, May 2025, doi: 10.1016/j.est.2025.116146.

[12] B. Ahmed, M. B. Tahir, A. Ali, and M. Sagir, "DFT insights on structural, electronic, optical and mechanical properties of double perovskites $X_2FeH_6$ (X = Ca and Sr) for hydrogen-storage applications," Int. J. Hydrog. Energy, vol. 50, pp. 316–323, Jan. 2024, doi: 10.1016/j.ijhydene.2023.10.237.

[13] T. Tang and Y. Tang, "Prediction of comprehensive properties and their hydrogen performance of $Mg_2XH_6$ (X=Mn, Fe, Co, Ni) perovskite hydrides based on first principles," Ceram. Int., vol. 50, no. 24, Part A, pp. 52270–52283, Dec. 2024, doi: 10.1016/j.ceramint.2024.10.078.

[14] Y.-L. Tao and Q.-J. Liu, "Investigation on the mechanism of electronic structure and superconductivity of cubic $X_2BH_6$ at ambient pressure," Mater. Today Phys., vol. 54, p. 101725, May 2025, doi: 10.1016/j.mtphys.2025.101725.

[15] S. Bahhar, A. Tahiri, M. Idiri, R. Touti, A. Jabar, and M. Naji, "Impact of pressure on quaternary Heusler alloy LiScNiGe for optoelectronic application," Mater. Sci. Semicond. Process., vol. 193, p. 109497, 2025.

[16] Z. El Fatouaki, A. Tahiri, A. Jabar, and M. Idiri, "First-principles study on the physical properties of double perovskites $LiX_3H_8$ (X = Ni and Mn) for hydrogen storage," J. Phys. Chem. Solids, vol. 206, p. 112867, Nov. 2025, doi: 10.1016/j.jpcs.2025.112867.

[17] Z. El Fatouaki, E. M. Hrida, A. Tahiri, A. Jabar, and M. Idiri, "Comprehensive first-principles and AIMD study of alkali metal $LiX_3H_8$ (X = Fe, Cr) hydrides for hydrogen storage applications," Int. J. Hydrog. Energy, vol. 163, p. 150791, Sept. 2025, doi: 10.1016/j.ijhydene.2025.150791.

[18] E. H. Akarchaou, R. Touti, A. El Mekkaouy, Y. Didi, A. Tahiri, and S. Chtita, "Computational analysis of X2MgTiH6 (X =Li, Na, and K) double perovskite hydride materials for hydrogen storage applications," Int. J. Hydrog. Energy, vol. 161, p. 150644, Aug. 2025, doi: 10.1016/j.ijhydene.2025.150644.

[19] R. Marwat et al., "Density Functional Theory (DFT) calculations for the adsorptive voltammetric determination of Meloxicam using a paste electrode made of Functionalized Carbon nanotubes," J. Indian Chem. Soc., vol. 102, no. 9, p. 101951, Sept. 2025, doi: 10.1016/j.jics.2025.101951.

[20] J. P. Perdew, K. Burke, and M. Ernzerhof, "Generalized gradient approximation made simple," Phys. Rev. Lett., vol. 77, no. 18, p. 3865, 1996.

[21] M. Ernzerhof and G. E. Scuseria, "Assessment of the Perdew–Burke–Ernzerhof exchange-correlation functional," J. Chem. Phys., vol. 110, no. 11, pp. 5029–5036, 1999.

[22] A. Alase, E. Cobanera, G. Ortiz, and L. Viola, "Generalization of Bloch's theorem for arbitrary boundary conditions: Theory," Phys. Rev. B, vol. 96, no. 19, p. 195133, 2017.

[23] J. Lv, S. Deng, and Z. Wan, "An efficient single-parameter scaling memoryless Broyden-Fletcher-Goldfarb-Shanno algorithm for solving large scale unconstrained optimization problems," IEEE Access, vol. 8, pp. 85664–85674, 2020.

[24] S. Y. Haffert, "The spectrally modulated self-coherent camera (SM-SCC): Increasing throughput for focal-plane wavefront sensing," Astron. Astrophys., vol. 659, p. A51, 2022.

[25] F. Elhamra, M. Rougab, and A. Gueddouh, "Theoretical investigation of transition metal (Cr, Fe)-Doped AlN in a rocksalt structure: A DFT study on physical properties," J. Phys. Chem. Solids, vol. 197, p. 112442, Feb. 2025, doi: 10.1016/j.jpcs.2024.112442.

[26] A. Ayyaz et al., "DFT investigation of thermodynamic, electronic, optical, and mechanical properties of $XLiH_3$ (X= Mg, Ca, Sr, and Ba) hydrides for hydrogen storage and energy harvesting," Mater. Sci. Semicond. Process., vol. 186, p. 109020, 2025.

[27] L. Ruihan, H. Feng, X. Ting, L. Yongzhi, Z. Xin, and Z. Jiaqi, "Progress in the application of first principles to hydrogen storage materials," Int. J. Hydrog. Energy, vol. 56, pp. 1079–1091, Feb. 2024, doi: 10.1016/j.ijhydene.2023.12.259.

[28] M. Jawad et al., "Unveiling the essential physical properties of indium-based thermodynamically stable delafossites $XInO_2$ (X= Na, K) as an energy harvesting material: a systematic first-principles study," Chem. Pap., pp. 1–17, 2025.

[29] Y. Zhu et al., "Explosion flame propagation and characteristics of coordination hydride hydrogen storage materials under concentration effects: A case study of $LiAlH_4$," Powder Technol., p. 121527, 2025.

[30] H. H. Raza, M. Naeem, H. S. Ali, A. Parveen, and A. M. Al-Enizi, "First-principles investigation of $BX_3H_9$ (X= Ca, Sc, Ti) hydrides: Structural, electronic, phonon, and hydrogen storage properties," J. Phys. Chem. Solids, p. 112800, 2025.

[31] Z. Abbas, D. Hussain, A. Alqahtani, and A. Parveen, "First-principles quantum analysis of physical and hydrogen storage properties of $XTi_3H_9$ (X= Rb, Cs and Fr) hydrides for hydrogen storage and optoelectronic applications," Int. J. Hydrog. Energy, vol. 151, p. 150218, 2025.

[32] C. Zenghua and M. Chunlan, "Revisiting the configurations of hydrogen impurities in $SrTiO_3$: Insights from first-principles local vibration mode calculations," ArXiv Prepr. ArXiv250705752, 2025.

[33] S. H. Mirza, N. H. Malik, C.-H. Zhan, and M. Jawad, "Enlightening the hydrogen storage potential of dynamically stable $Mg_2TmH_6$ (Tm = Co, Ni) perovskite hydrides: Prospects for hydrogen storage applications," Int. J. Hydrog. Energy, vol. 160, p. 150562, 2025.

[34] M. E. Tuckerman, "Ab initio molecular dynamics: basic concepts, current trends and novelapplications," J. Phys. Condens. Matter, vol. 14, no. 50, p. R1297, 2002.

[35] P.-C. Wei et al., "Thermodynamic routes to ultralow thermal conductivity and high thermoelectric performance," Adv. Mater., vol. 32, no. 12, p. 1906457, 2020.

[36] A. Kostopoulou, E. Kymakis, and E. Stratakis, "Perovskite nanostructures for photovoltaic and energy storage devices," J. Mater. Chem. A, vol. 6, no. 21, pp. 9765–9798, 2018.

[37] M. El Akkel and H. Ez-Zahraouy, "Novel double hydride perovskites $Li_2TiF_{6-x}H_x$ as efficient materials for solid-state hydrogen storage: DFT insights," Int. J. Hydrog. Energy, vol. 101, pp. 1406–1420, 2025.

[38] E. M. Hrida, Z. El Fatouaki, O. Zedouh, A. Tahiri, and M. Idiri, "Investigation of LiScNiSi Heusler alloy's physical characteristics under pressure: Use in optoelectronics with the HSE06 hybrid function," Solid State Commun., vol. 404, p. 116094, Oct. 2025, doi: 10.1016/j.ssc.2025.116094.

[39] Q. Dai, T.-Y. Tang, Z.-Q. Chen, Y. Wang, and Y.-L. Tang, "A DFT study to investigate of $K_2LiXH_6$ (X= Al, Ga, In) perovskite hydrides for hydrogen storage application," Int. J. Hydrog. Energy, vol. 101, pp. 295–302, 2025.

[40] M. Archi, O. Bajjou, and B. Elhadadi, "A comparative ab initio analysis of the stability, electronic, thermodynamic, mechanical, and hydrogen storage properties of $SrZnH_3$ and $SrLiH_3$ perovskite hydrides through DFT and AIMD Approaches," Int. J. Hydrog. Energy, vol. 105, pp. 759–770, 2025.

[41] A. Ayyaz et al., "Investigation of hydrogen storage and energy harvesting potential of double perovskite hydrides $A_2LiCuH_6$ (A= Be/Mg/Ca/Sr): A DFT approach," Int. J. Hydrog. Energy, vol. 102, pp. 1329–1339, 2025.

[42] A. Ayyaz et al., "DFT investigation of thermodynamic, electronic, optical, and mechanical properties of $XLiH_3$ (X= Mg, Ca, Sr, and Ba) hydrides for hydrogen storage and energy harvesting," Mater. Sci. Semicond. Process., vol. 186, p. 109020, 2025.

[43] Y. Selmani and L. Bahmad, "Insights into the physical properties of $NaGeH_3$ perovskite hydride for hydrogen storage applications: A first-principles study," J. Phys. Chem. Solids, vol. 208, p. 113089, Jan. 2026, doi: 10.1016/j.jpcs.2025.113089.

[44] M. El Akkel, M. Achqraoui, N. Bekkioui, and H. Ez-Zahraouy, "Strain engineering, ionic substitution and co-substitution: Pathways to enhanced hydrogen storage performance of $KMgH_3$," Chem. Phys., vol. 600, p. 112910, Jan. 2026, doi: 10.1016/j.chemphys.2025.112910.

[45] K. Aafi, Z. El Fatouaki, A. Jabar, A. Tahiri, and M. Idiri, "New alkali metal compounds $XCo_3H_8$ (X = Li, Na and K) for hydrogen storage technology," J. Phys. Chem. Solids, vol. 208, p. 113064, Jan. 2026, doi: 10.1016/j.jpcs.2025.113064.

[46] B. Ahmed, M. B. Tahir, A. Dahshan, and M. Sagir, "Advanced computational screening of $X_2CaH_4$ (X = Rb and Cs) for hydrogen storage applications," Int. J. Hydrog. Energy, vol. 89, pp. 48–55, Nov. 2024, doi: 10.1016/j.ijhydene.2024.09.321.

[47] M. M. Parvaiz, A. Khalil, H. I. Elsaeedy, M. B. Tahir, S. Ayub, and Z. Ullah, "Extensive screening of novel $BaXH_3$ (X = V, Cr, Co, Ni, Cu, and Zn) perovskites for physical properties and hydrogen storage application: A DFT study," Int. J. Hydrog. Energy, vol. 87, pp. 1056–1073, Oct. 2024, doi: 10.1016/j.ijhydene.2024.09.113.

[48] M. Usman, N. Bibi, S. Rahman, M. Awais Rehman, and S. Ahmad, "A DFT study to investigate $BeXH_3$ (X = Ti, Zr) hydride perovskites for hydrogen storage application," Comput. Theor. Chem., vol. 1240, p. 114820, Oct. 2024, doi: 10.1016/j.comptc.2024.114820.

[49] Z. Qian et al., "Unveiling and boosting the catalytic activity of $LaBO_3$ (B= Mn, Fe, Co, Ni) perovskites: d-p band center distance as a descriptor," Appl. Catal. B Environ. Energy, vol. 381, p. 125864, Feb. 2026, doi: 10.1016/j.apcatb.2025.125864.

[50] A. El-Khouly et al., "Mechanical and crystallographic texture features of half-Heusler TiCoSb and based alloys," J. Alloys Compd., vol. 1037, p. 182539, Aug. 2025, doi: 10.1016/j.jallcom.2025.182539.

[51] R. Gaillac, P. Pullumbi, and F.-X. Coudert, "ELATE: an open-source ne application for analysis and visualization of elastic tensors," J. Phys. Condens. Matter, vol. 28, no. 27, p. 275201, 2016.

[52] M. L. Cohen and J. R. Chelikowsky, Electronic structure and optical properties of semiconductors, vol. 75. Springer Science & Business Media, 2012.

[53] T. Y. Ahmed, S. B. Aziz, and E. M. A. Dannoun, "New photocatalytic materials based on alumina with reduced band gap: A DFT approach to study the band structure and optical properties," Heliyon, vol. 10, no. 5, Mar. 2024, doi: 10.1016/j.heliyon.2024.e27029.

[54] D. Nematov, "Analysis of the Optical Properties and Electronic Structure of Semiconductors of the $Cu_2NiXS_4$ (X = Si, Ge, Sn) Family as New Promising Materials for Optoelectronic Devices," J. Opt. Photonics Res., vol. 1, no. 2, pp. 91–97, Jan. 2024, doi: 10.47852/bonviewJOPR42021819.

[55] E. Darkaoui et al., "Investigation of $Ba_2SmOsO_6$ double perovskite oxide for spintronic and linear optoelectronic applications using DFT and DFT+U methods," J. Magn. Magn. Mater., vol. 622, p. 172980, June 2025, doi: 10.1016/j.jmmm.2025.172980.

[56] M. Kurban, C. Polat, E. Serpedin, and H. Kurban, "Enhancing the electronic properties of $TiO_2$ nanoparticles through carbon doping: An integrated DFTB and computer vision approach," Comput. Mater. Sci., vol. 244, p. 113248, Sept. 2024, doi: 10.1016/j.commatsci.2024.113248.

[57] Z. Lv et al., "Asymmetric high-coordination Co-NSP single-atom catalysts with tailored d-p-orbital electron structure for efficient bifunctional catalyst of rechargeable Zn-air battery cathodes," Appl. Catal. B Environ. Energy, vol. 365, p. 124889, May 2025, doi: 10.1016/j.apcatb.2024.124889.

[58] Y. Xu, Y. Zhou, Y. Li, Y. Hao, P. Wu, and Z. Ding, "Magnesium-Based Hydrogen Storage Alloys: Advances, Strategies, and Future Outlook for Clean Energy Applications," Molecules, vol. 29, no. 11, p. 2525, Jan. 2024, doi: 10.3390/molecules29112525.

[59] H. N. Yoğurtçu and C. C. Ersanlı, "Structural Parameters, NLO, HOMO, LUMO, MEP, Chemical Reactivity Descriptors, Mulliken-NPA, Thermodynamic Functions, Hirshfeld Surface Analysis and Molecular Docking of 1,3-Bis(4-methylphenyl)triazine," Int. Sci. Vocat. Stud. J., vol. 9, no. 1, pp. 130–144, June 2025, doi: 10.47897/bilmes.1697802.

[60] M. Mubeen Parvaiz, A. Khalil, M. B. Tahir, S. Ayub, T. E. Ali, and H. Tariq Masood, "A DFT investigation on structural, electronic, magnetic, optical, elastic and hydrogen storage properties of Ru-based hydride-perovskites $XRuH_3$ (X = Cr, V, Ni)," RSC Adv., vol. 14, no. 12, pp. 8385–8396, 2024, doi: 10.1039/D4RA00204K.

[61] T. Yaseen Ahmed, S. B. Aziz, and E. M. A. Dannoun, "Role of outer shell electron-nuclear distant of transition metal atoms (TMA) on band gap reduction and optical properties of $TiO_2$ semiconductor," Results Eng., vol. 23, p. 102479, Sept. 2024, doi: 10.1016/j.rineng.2024.102479.

[62] P. Xie, W. Wang, and Y. Kivshar, "Resonant light–matter interaction with epsilon-near-zero photonic structures," Appl. Phys. Rev., vol. 12, no. 2, p. 021307, Apr. 2025, doi: 10.1063/5.0252120.

[63] A. Ou-khouya, I. Ait Brahim, H. Ez-Zahraouy, A. Houba, H. Mes-Adi, and M. Tahiri, "First-principles calculations to investigate impact of doping by chalcogen elements on the electronic, structural, and optical properties of $SrTiO_3$ compounds," Chem. Phys., vol. 581, p. 112253, May 2024, doi: 10.1016/j.chemphys.2024.112253.

[64] R. Song et al., "Exploring the structural, physical and hydrogen storage properties of Cr-based perovskites $YCrH_3$ (Y = Ca, Sr, Ba) for hydrogen storage applications," Ceram. Int., vol. 50, no. 20, Part B, pp. 39739–39747, Oct. 2024, doi: 10.1016/j.ceramint.2024.07.353.

[65] M. H. Abbas, H. Ibrahim, A. Hashim, and A. Hadi, "Fabrication and Tailoring Structural, Optical, and Dielectric Properties of $PS/CoFe_2O_4$ Nanocomposites Films for Nanoelectronics and Optics Applications," Trans. Electr. Electron. Mater., vol. 25, no. 4, pp. 449–457, Aug. 2024, doi: 10.1007/s42341-024-00524-5.

[66] N. Al-Zoubi, A. Almahmoud, A. Almahmoud, and A. Obeidat, "Theoretical assessment of a novel $NaXH_3$ and $KXH_3$ (X = Tc, Ru and Rh) perovskite hydrides for hydrogen storage," Int. J. Hydrog. Energy, vol. 93, pp. 822–831, Dec. 2024, doi: 10.1016/j.ijhydene.2024.11.020.

[67] A. M. Alsuhaibani et al., "First-principle Insight into Structural, Electronic, Optical and Elastic Properties of $AgXF_3$ (Cr, Zr) Halide Perovskite Materials for Application of Reflective Coating," J. Inorg. Organomet. Polym. Mater., vol. 34, no. 8, pp. 3613–3622, Aug. 2024, doi: 10.1007/s10904-024-03035-1.

[68] F. A. Nelson et al., "Chemical effect of alkaline-earth metals (Be, Mg, Ca) substitution of $BFe_2XH$ hydride perovskites for applications as hydrogen storage materials: A DFT perspective," Int. J. Hydrog. Energy, vol. 79, pp. 1191–1200, Aug. 2024, doi: 10.1016/j.ijhydene.2024.07.063.

[69] H. Murtaza, S. M. H. Qaid, H. M. Ghaithan, A. A. Ali Ahmed, and J. Munir, "The excellent performance of $AClH_3$ (A=Rb, Cs, K) perovskite hydrides for hydrogen storage applications," Renew. Energy, vol. 252, p. 123491, Oct. 2025, doi: 10.1016/j.renene.2025.123491.

[70] M. Usman et al., "Hydrogen storage application of Zn-based hydride-perovskites: a computational insight," Opt. Quantum Electron., vol. 56, no. 9, p. 1478, Sept. 2024, doi: 10.1007/s11082-024-07399-z.

[71] K. Ikeda, Y. Kogure, Y. Nakamori, and S. Orimo, "Formation region and hydrogen storage abilities of perovskite-type hydrides," Prog. Solid State Chem., vol. 35, no. 2, pp. 329–337, Jan. 2007, doi: 10.1016/j.progsolidstchem.2007.01.005.

[72] A. Siddique et al., "Structures and hydrogen storage properties of AeVH$_3$ (Ae = Be, Mg, Ca, Sr) perovskite hydrides by DFT calculations," Int. J. Hydrog. Energy, vol. 48, no. 63, pp. 24401–24411, July 2023, doi: 10.1016/j.ijhydene.2023.03.139.

[73] R. Balderas-Xicohténcatl, M. Schlichtenmayer, and M. Hirscher, "Volumetric Hydrogen Storage Capacity in Metal–Organic Frameworks," Energy Technol., vol. 6, no. 3, pp. 578–582, 2018, doi: 10.1002/ente.201700636.

[74] C. Liu, F. Li, L.-P. Ma, and H.-M. Cheng, "Advanced Materials for Energy Storage," Adv. Mater., vol. 22, no. 8, pp. E28–E62, 2010, doi: 10.1002/adma.200903328.

[75] H. Wenfeng et al., "Kinetic characteristics and optimization of hydrogen absorption in carbon-based materials," Renew. Energy, vol. 256, p. 124238, Jan. 2026, doi: 10.1016/j.renene.2025.124238.

[76] A. Assila et al., "The role of Al substitution in Na$_3$AlH$_6$ hydrides: Structural and thermodynamic insights for hydrogen storage technologies," J. Power Sources, vol. 634, p. 236502, Apr. 2025, doi: 10.1016/j.jpowsour.2025.236502.

[77] P. Girhe, D. P. Barai, B. A. Bhanvase, and S. H. Gharat, "A Review on Functional Materials for Hydrogen Storage," Energy Storage, vol. 7, no. 5, p. e70218, 2025, doi: 10.1002/est2.70218.